\documentclass[a4paper]{article}
\usepackage[left=1in,top=1in,right=1in,bottom=1in,nohead]{geometry}
\usepackage{authblk}
\usepackage{blindtext}
\usepackage[T1]{fontenc}
\usepackage[utf8]{inputenc}
\usepackage{lmodern}

\usepackage[english]{babel}
\usepackage{csquotes}
\usepackage[english]{babel}
\usepackage{amsfonts}
\usepackage{amssymb}
\usepackage{amsmath}
\usepackage{amsthm}
\usepackage{eucal}
\usepackage{mathrsfs}
\allowdisplaybreaks[0]
\usepackage{bm}
\usepackage{graphicx}
\usepackage{caption}
\usepackage{subcaption}
\usepackage{algorithm,algorithmic}
\usepackage{framed}
\usepackage{diagbox}
\usepackage[titletoc,title]{appendix}
\usepackage{tikz}
\usepackage{color}
\usepackage{hyperref}
\usepackage{cleveref}
\usepackage{autonum}
\title{Overcomplete}
\date{}

\DeclareGraphicsExtensions{.eps,.pdf}

\usepackage{enumitem}
\setlist[enumerate]{leftmargin=.5in}
\setlist[itemize]{leftmargin=.5in}

\begin{document}
	
\title{Space-variant Generalized Gaussian Regularization for Image Restoration}

\author{A Lanza$^{\rm a}$, S. Morigi$^{\rm a}$, M. Pragliola$^{\rm a}$ and F. Sgallari$^{\rm a}$\\\vspace{6pt} $^{a}${\em{Department of Mathematics, University of Bologna, Piazza di Porta San Donato 5, Bologna, IT}}}
\maketitle

\begin{abstract}
We propose a new space-variant regularization term for variational image restoration based on the assumption that the
gradient magnitudes of the target image distribute locally according to a half-Generalized Gaussian distribution. 
This leads to a highly flexible regularizer characterized by two per-pixel free parameters, 
which are automatically estimated from the observed image.
The proposed regularizer is coupled with either the $L_2$ or the $L_1$ fidelity terms, in order to effectively deal with additive 
white Gaussian noise or impulsive noises such as, e.g, additive white Laplace and salt and pepper noise.
The restored image is efficiently computed by means of an iterative numerical algorithm based on the alternating direction method of multipliers.
Numerical examples indicate that the proposed regularizer holds the potential for achieving high quality restorations for a wide range 
of target images characterized by different gradient distributions and for the different types of noise considered.
\end{abstract}

\section{Introduction}
\label{sec:intro}
Image restoration refers to the recovery of a clean sharp image from a noisy, and potentially blurred, observation. 
In this paper, we consider the problem of restoring images corrupted by known blur and different types of noise.

We consider gray level images of size $\,d_1 \!{\times}\, d_2$, such that \mbox{$n \,{:=}\, d_1 d_2$} is the total number of pixels in the images.
The general model of the image degradation process under blur and noise corruptions can be written as
\begin{equation}
g \:\;{=}\;\: \mathcal{N}\left( K u \right) \: ,
\label{eq:GDM}
\end{equation}
where $u, g \in \mathbb{R}^{n}$ represent vectorized forms of the unknown clean image and of the observed corrupted image, respectively,
$K \in \mathbb{R}^{n \times n}$ is a known linear blurring operator and $\mathcal{N}(\,\cdot\,)$ denotes the noise corruption operator, 
which in most cases is of random nature.
Given $K$ and $g$, the goal of image restoration is to solve the ill-conditioned - or even singular, depending on $K$ - inverse problem
of recovering an as accurate as possible estimate $u^*$ of the unknown clean image $u$.

The class of \emph{variational methods} for image restoration relies on determining
restored images $u^*\in\mathbb{R}^n$ as the minimizers of suitable cost functionals $\mathcal{J}(u): \mathbb{R}^n \to \mathbb{R}$ such that, 
typically, restoration is casted as an optimization problem of the form
%
\begin{equation}
u^* \:\;{\leftarrow}\;\: \arg \min_{u \in \mathbb{R}^n}
\mathcal{J}(u) \, , \qquad
\mathcal{J}(u) \;{:=}\; \mathcal{R}(u) \;{+}\; \mu \, \mathcal{F}(u;g)
\, ,
\label{eq:GVM}
\end{equation}
where the functionals $\mathcal{R}(u)$ and $\mathcal{F}(u;g)$, commonly referred as the \emph{regularization} and the \emph{fidelity} term,
encode prior information on the clean image $u$ and on the observation model (\ref{eq:GDM}), respectively,
with the so-called regularization parameter $\mu > 0$ controlling the trade-off between the two terms.

The functional form of the fidelity term is strictly connected with the characteristics of the noise corruption.
In this paper, we are interested in three important types of noise, namely the additive (zero-mean) white Gaussian noise (AWGN) which typically appears, e.g., in Magnetic Resonance Tomography, the additive (zero-mean) white Laplace noise (AWLN) and the impulsive salt and pepper noise (SPN) usually due to transmission errors or malfunctioning pixel elements in camera sensors.
Denoting by $\Omega := \{1,\ldots,n\}$ the set of all
pixel positions in the vectorized images, for these three kinds of noise the general degradation model in (\ref{eq:GDM}) reads as
\begin{equation}
\begin{array}{ccc}
\mathrm{AWGN}$\;$\mathrm{and}$\;$\mathrm{AWLN:} & \quad\;\; & \mathrm{SPN:} \vspace{0.1cm} \\
g_i \:\;{=}\;\: (K u)_i \;{+}\; n_i \;\;\: \forall \, i \in \Omega \, , & &
g_{i} \:\;{=}\;\: \left\{
\begin{array}{ll}
(K u)_i \;\; & \mathrm{for} \;\;\: i \in \Omega_0 \subseteq \Omega\\
n_i \in \{0,1\}          & \mathrm{for} \;\;\: i \in {\Omega_1} :=\Omega \setminus \Omega_0
\end{array} \right. \, .
\end{array}
\label{eq:SDM}
\end{equation}
%
%
%
For what concerns AWGN and AWLN, the additive corruptions $n_i \in \mathbb{R}$, $i \in \Omega$, represent independent realizations from the same univariate Gaussian and Laplace distribution with zero mean and standard deviation $\sigma$, respectively. In the case of SPN, only a subset $\Omega_1$ of
the pixels is corrupted by noise, whereas the complementary subset $\Omega_0$ is noise-free. In particular, the corrupted pixels can take only the two possible extreme values $\{0,1\}$ (we assume that images have range $[0,1]$), with the same probability.
The subset $\Omega_1$ is known in some applications \cite{Serena17} or it could be estimated \cite{mila}. 
As the zero-mean AWGN and AWLN are fully characterized
from a probabilistic point of view by the unique scalar parameter $\,\sigma$, SPN is characterized by the parameter
$\gamma \in [0,1]$ which represents the probability for a pixel to be noise-corrupted. 

It is well known that AWGN and the impulsive AWLN and SPN
are suitably dealt with by the so-called L$_2$ and L$_1$ fidelity terms, which are related to the $\ell_2$
and $\ell_1$ norm of the residue image, respectively \cite{ADMTV}; in formulas:
\begin{equation}
\mathcal{F}(u;g) \,\;{=}\;\, \mathrm{L}_q(u;g) \,\;{:=}\;\, \frac{1}{q} \, \| K u - g \|_q^q ,
\quad q \in \{1,2\} \, .
\label{eq:Lq}
\end{equation}

For what regards the regularization term in (\ref{eq:GVM}), a very popular choice
is represented by the Total Variation (TV) semi-norm \cite{ROF}, that is
\begin{equation}
\mathcal{R}(u) \:\;{=}\;\: \mathrm{TV}(u) \,\;{:=}\; \sum_{i=1}^{n}  \| (\nabla u)_{i} \|_2 \, ,
\label{eq:TV}
\end{equation}
where $(\nabla u)_i := \big( (D_h u)_i , (D_v u)_i \big)^T \in \mathbb{R}^2$ denotes the discrete gradient of image $u$ at pixel $i$,
with $D_h,D_v \in \mathbb{R}^{n \times n}$ linear operators representing finite difference discretizations of the first-order
horizontal and vertical partial derivatives, respectively.
Popularity of TV regularizer for image restoration is mainly due to two facts, namely it is convex and allows for restored images with sharp, neat edges. By substituting the TV regularizer (\ref{eq:TV}) and the L$_2$ or L$_1$ fidelity terms (\ref{eq:Lq}) for $\mathcal{R}$ and $\mathcal{F}$ in (\ref{eq:GVM}),
respectively, one obtains the so-called TV-L$_2$ \cite{ROF} - or ROF - and TV-L$_1$ \cite{tvl1} restoration models, which reads as
\begin{equation}
u^* \:\;{\leftarrow}\;\: \arg \min_{u \in \mathbb{R}^n}
\left\{ \,
\mathrm{TV}(u) \,\;{+}\;\, \mu \, \mathrm{L}_q(u;g)
\, \right\} ,
\quad q \in \{1,2\} \, .
\label{eq:TVLq}
\end{equation}
The TV-L$_2$ and TV-L$_1$ models in (\ref{eq:TVLq}) are non-smooth convex and allow to obtain good quality restorations of images corrupted by AWGN and AWLN/SPN, respectively, such that they are regarded as sort of baseline models.
%
%

The contribution of this paper consists of a new space-variant regularization term which, coupled with the $L_2$ or $L_1$ fidelity terms, 
gives rise to a generalization of the models in (\ref{eq:TVLq}) of the form
\begin{equation}
u^* \:\;{\leftarrow}\;\: \arg \min_{u \in \mathbb{R}^n}
\left\{ \,
\mathrm{TV}_{p,\alpha}^{\mathrm{sv}}(u) \,\;{+}\;\, \mu \, \mathrm{L}_q(u;g)
\, \right\} ,
\quad q \in \{1,2\} \, ,
\label{eq:PMa}
\end{equation}
where the new space-variant TV$_{p,\alpha}^{\mathrm{sv}}$ regularizer 
is defined by
\begin{equation}
\mathrm{TV}_{p,\alpha}^{\mathrm{sv}}(u) \,\;{:=}\; \sum_{i=1}^{n}  \alpha_i \| (\nabla u)_{i} \|_2^{p_i} ,
\quad \alpha_i \:{\in}\; ]0,+\infty[, \;\: p_i \:{\in}\; ]0,2] \;\; \forall \, i \in \Omega \, .
\label{eq:PMb}
\end{equation}
%

%
%
The proposed regularizer in (\ref{eq:PMb}) is highly flexible as it is characterized by two per-pixel free parameters 
$p_i$, $\alpha_i$, such that local, space-variant properties of the target clean image $u$ can be potentially addressed.
The usefulness of such a great flexibility in the proposed regularizer is however conditioned to the existence of effective
procedures for the automatic estimation of the $p_i$ and $\alpha_i$ parameters. 
Hence, in this paper we also propose a suitable method for the automatic estimation of such parameters from the observed image 
partially based on the statistical inference technique described in~\cite{shape2}. 

As outlined in Section \ref{sec:map}, the rationale of our proposal is that the distribution of the gradient magnitudes of the unknown clean image is space-variant and it is well modeled 
locally by a two-parameters Generalized Gaussian distribution. 
As highlighted in \cite{tvpl2}, the TV regularizer in (\ref{eq:TV}) comes from implicitly assuming a space-invariant (that is, frame-based), one-parameter half-Laplacian distribution for the gradient magnitudes. Based on the observation that such a distribution is not sufficiently flexible, the authors  
in~\cite{tvpl2} proposed a generalization of the TV regularizer, referred to as TV$_p$, which relies on a space-invariant, two-parameters half-Generalized Gaussian distribution with $p$ denoting the additional parameter, called the \emph{shape} parameter. Finally, the very recently proposed TV$_{p}^{\mathrm{sv}}$ regularizer \cite{vip} further generalizes the TV$_p$ by assuming a space-variant shape parameter in the two-parameters half-Generalized Gaussian distribution. The authors in \cite{vip} demonstrated experimentally how using a local, space-variant model holds the potential for achieving higher quality restorations than using a global, space-invariant model.

The numerical solution of the two proposed variational models (\ref{eq:PMa})--(\ref{eq:PMb}) are obtained by means of an efficient iterative minimization 
algorithm based on the Alternating Direction Method of Multipliers (ADMM) strategy \cite{BOYD_ADMM}.

The paper is organized as follows. The derivation of the models \eqref{eq:PMa} and a more detailed motivation of their introduction are proposed in Section \ref{sec:map}. In Section \ref{sec:pi} we briefly outline the statistical inference procedure used 
for automatically estimating the $p_i$ and $\alpha_i$ parameters. The ADDM-based minimization algorithm is
described in detail in Section \ref{sec:admm}. Some meaningful numerical results are reported
in Section \ref{sec:nr} and, finally, conclusions are drawn in Section \ref{sec:conc}.

\section{Motivation via MAP estimator}
\label{sec:map}
The Maximum A Posteriori (MAP) Estimation approach \cite{bov} relies on the maximization of the posterior probability density function Pr$(u|g;K)$ associated to the clean unknown image $u$: 
\begin{equation}
u^* \:\;{\leftarrow}\;\: \arg \max_{u \in \mathbb{R}^n}\;
\text{Pr}(u|g;K).
\label{eq:map1}
\end{equation}
Relying on the Bayes' formula, and dropping the \emph{evidence} term Pr$(g)$, this is equivalent to maximize the product of the \emph{prior} Pr$(u)$ and the \emph{likelihood} Pr$(g|u;K)$ probability density functions.
By taking the negative logarithm of this product, problem \eqref{eq:map1} can be reformulated as follows:
\begin{equation}
u^* \:\;{\leftarrow}\;\: \arg \min_{u \in \mathbb{R}^n} \;
\left\{ \,-\log \text{Pr}(g|u;K)-\log \text{Pr}(u)
\, \right\}.
\label{eq:map2}
\end{equation}
At first, we focus on the setting of the prior. A common choice is to model the unknown image $u$ as a Markov Random Field (MRF) such that the image can be characterized by its Gibbs prior distribution, whose general form is:
\begin{equation}
\text{Pr}(u)= \frac{1}{Z} \prod_{i = 1}^{n} \text{exp}\,(\,-\alpha \, V_{c_{i}}(u)\,)=  \frac{1}{Z} \text{exp}\,\bigg(\,-\alpha\,\sum_{i = 1}^{n}  V_{c_{i}}(u)\,\bigg),
\label{eq:mrf}
\end{equation}
where $\alpha>0$ is the MRF parameter, $\{c_{i}\}_{i=1}^{n}$ is the set of all cliques (a clique is a set
of neighboring pixels) for the MRF, $V_{c_{i}}$ is the potential function defined on the clique $c_{i}$
and $Z$ is the partition function, that is a function not depending on $u$ which allows for the normalization of the prior. 
\\Choosing as potential function at the generic $i$-th pixel the magnitude of the discrete gradient at the same pixel, i.e. $V_{c_{i}}=\lVert (\nabla u)_{i} \rVert_{2}$, the Gibbs prior in \eqref{eq:mrf} reduces to the popular TV prior:
\begin{equation}
\text{Pr}(u)= \frac{1}{Z} \,\text{exp}\,\bigg(\,-\alpha\,\sum_{i = 1}^{n} \lVert (\nabla u)_{i} \rVert_{2}  \,\bigg)=\frac{1}{Z} \, \text{exp}\bigg(-\alpha \, \text{TV} (u)\,\bigg) ,
\label{eq:tv}
\end{equation} 
where $Z$ is the normalization constant not depending on $u$.
The adoption of a TV prior can be further interpreted as assuming that the $\ell_{2}$ norm of the gradient
at any pixel of the unknown clean image, $\lVert (\nabla u)_{i} \rVert_{2}$, follows a space-invariant half-Laplacian (or exponential)
distribution:

$$\text{Pr} (x;\alpha) \:\;{=}\;\: \left\{
\begin{array}{ll}
\alpha\, \exp\,(\,-\alpha \,x\,) & \mathrm{for} \;\;\: x \geq 0\\
&\\
0          & \mathrm{for} \;\;\: x<0
\end{array} \right. \, .$$

\begin{figure}[tbh]
	\center
	\begin{tabular}{ccc}
		\includegraphics[width=1.5in]{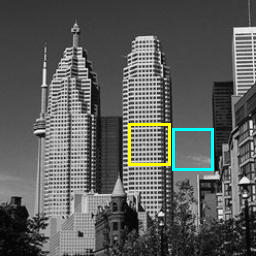} &
		\includegraphics[width=2.1in]{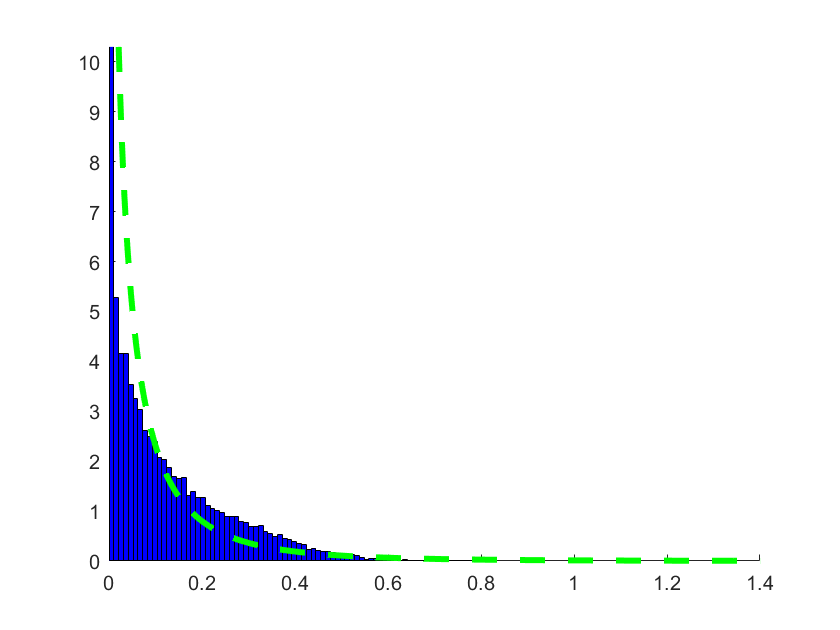} &
		\includegraphics[width=2.1in]{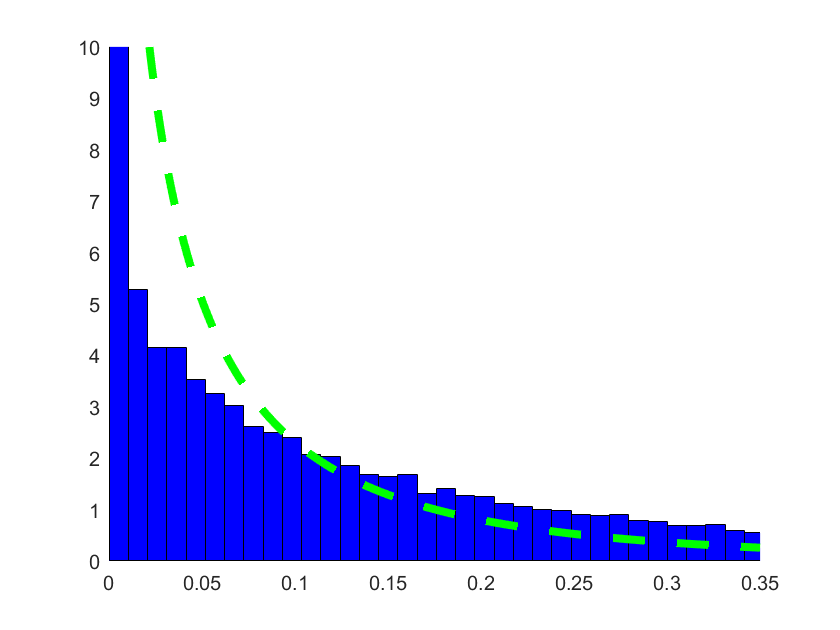}\\
		(a) test image&(b) global histogram &(c) zoom of (b)\\
		\includegraphics[width=1.5in]{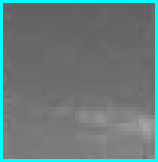} &
		\includegraphics[width=2.1in]{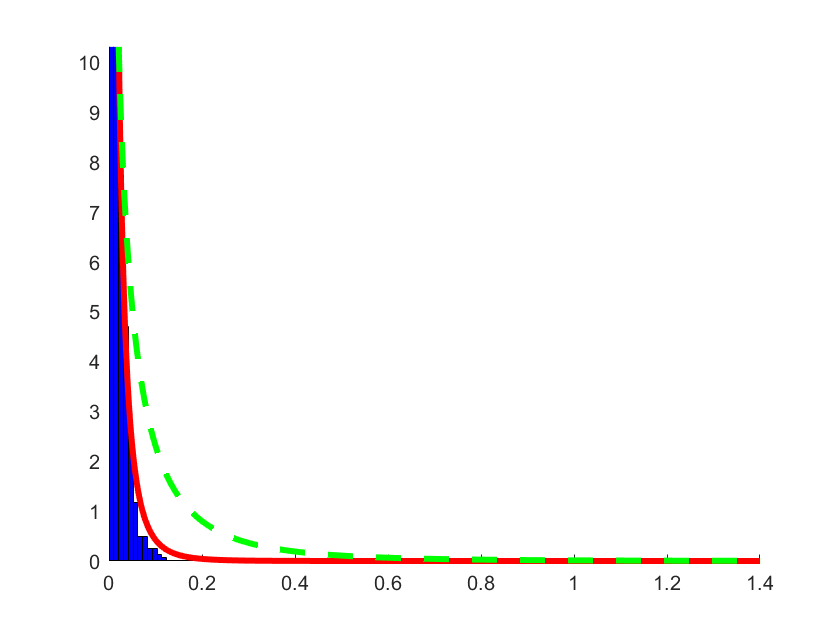} & 
		\includegraphics[width=2.1in]{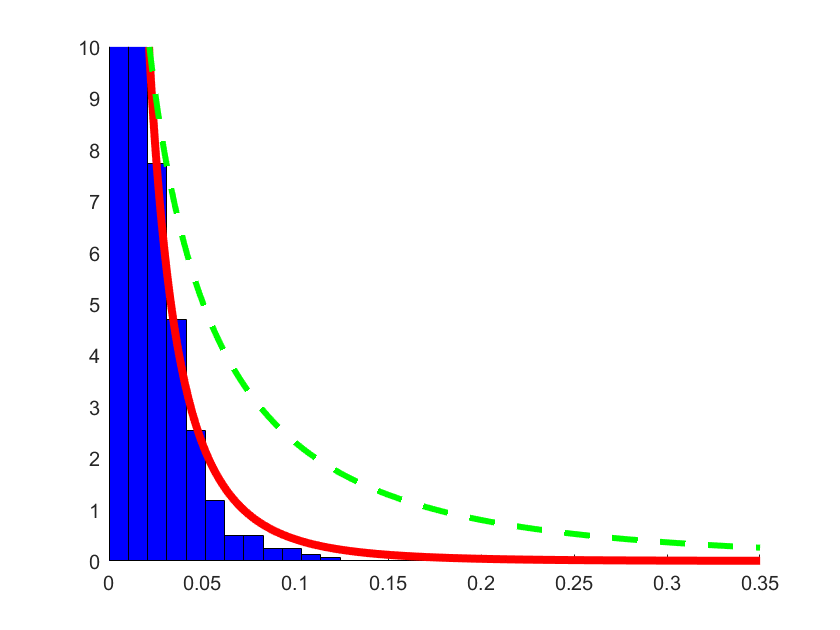}\\
		(d) smooth region &(e) local histogram for (d)&(f) zoom of (e)\\ 
		
		\includegraphics[width=1.5in]{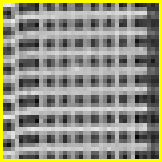} &
		\includegraphics[width=2.1in]{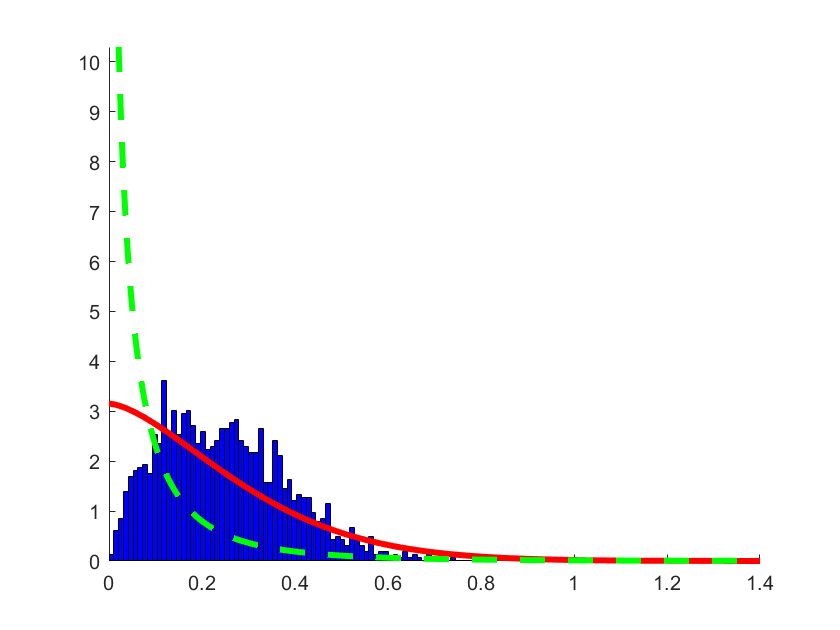} & 
		\includegraphics[width=2.1in]{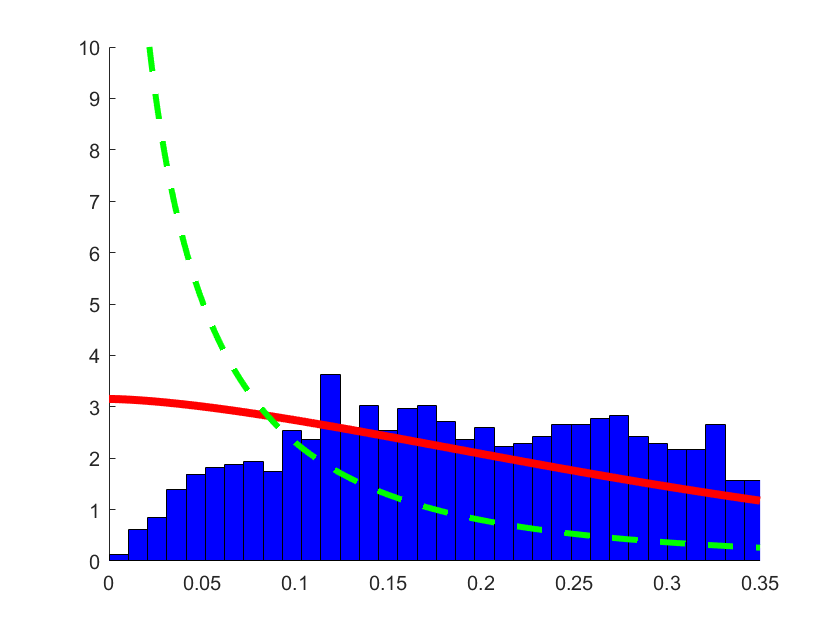}\\
		(g) texture region&(h) local histogram for (g)&(i) zoom of (h)\\ 
		\end{tabular}		
	\caption{Gradient magnitudes histograms on the whole test image, on a smooth region and on a texture region.}
	\label{fig:hist}
\end{figure}

In \cite{tvpl2}, a deep investigation about the effect of replacing the half-Laplacian distribution with the more flexible half-Generalized Gaussian distribution
\begin{equation}
\text{Pr}(x;p,\alpha) \:\;{=}\;\: \left\{
\begin{array}{ll}
\frac{\alpha p}{\Gamma(1/p)} \exp(-(\alpha x)^{p}) & \mathrm{for} \;\;\: x \geq 0\\
&\\
0          & \mathrm{for} \;\;\: x<0
\end{array} \right. \, 
\label{eq:hgg}
\end{equation}
has been carried out. The presence of a second parameter $p$   allows for a better approximation of the $\ell_{2}$ norm gradient distribution and leads to the introduction of the  TV$_{p}$ prior:
\begin{equation}
\text{Pr}(u)= \frac{1}{Z} \, \text{exp}\,\bigg(\,-\alpha\,\sum_{i = 1}^{n} \lVert (\nabla u)_{i} \rVert_{2}^{p}  \,\bigg)=\frac{1}{Z} \, \text{exp}\bigg(-\alpha \, \text{TV}_{p} (u)\,\bigg).
\label{eq:tvp}
\end{equation} 
In this paper, we propose a prior, consisting in a non-stationary (space-variant) Markov Random Field. The parameters $\alpha,p$ of the half-Generalized Gaussian distribution of the magnitude of the discrete gradients change as the clique $c_{i}$ changes. Therefore, the prior takes the following form:
\begin{equation}
\text{Pr}(u)= \frac{1}{Z} \,\text{exp}\,\bigg(\,-\,\sum_{i = 1}^{n} \alpha_{i} \lVert (\nabla u)_{i} \rVert_{2}^{p_{i}}  \,\bigg)=\frac{1}{Z} \, \text{exp}\bigg(-\, \text{TV}_{p,\alpha}^{\mathrm{sv}} (u)\,\bigg).
\label{eq:nsmrf}
\end{equation} 
The adoption of a space-variant approach is expected to be more flexible for the restoration of images presenting cliques with different properties, i.e. images in which texture, smooth, piecewise constant regions, and edges co-exist.
\\In order to justify the reason why a space-variant model should be adopted in general, we consider the test image \texttt{skyscraper} illustrated in Fig. \ref{fig:hist}(a). We selected two regions characterized by smooth and texture structures - see the cyan-bordered and the yellow-bordered boxes, respectively, in Figs. \ref{fig:hist}(d),\ref{fig:hist}(g). 
\\The histogram of the gradient magnitudes of the whole image is shown in Fig \ref{fig:hist}(b) and zoomed in Fig. \ref{fig:hist}(c). The superimposed green dashed lines, which have been reproduced in each sub-figure for comparison, represent the half-Generalized Gaussian distributions that best fit the histograms and whose parameters have been computed considering all the pixels of the test image. 
\\The histogram of the gradient magnitudes in the two bordered regions are shown in Figs. \ref{fig:hist}(e), \ref{fig:hist}(h) and zoomed in Figs. \ref{fig:hist}(f), \ref{fig:hist}(i). The red solid lines represent the half-Generalized Gaussian distributions that best fit the histograms and whose parameters have been computed considering only the pixels in the boxes.

It is worth noticing how the histograms of the gradient magnitudes in the two selected regions are very different from each other and also differ from the one of the whole test image. As a result, the red lines fit the histogram shapes in Figs. \ref{fig:hist}(e), \ref{fig:hist}(h) better than the green ones - see also the zooms in Figs. \ref{fig:hist}(f),\ref{fig:hist}(i). This is the benefit of the space-variant strategy, which is able to model space-variant image features.\\

\noindent Going back to the MAP inference formula \eqref{eq:map2}, in particular to the likelihood term Pr$(g|u;K)$, assuming the noise to be additive independent identically distributed, we have:
\begin{equation}
\text{Pr}(g|u;K)=\prod_{i=1}^{n}\,\text{Pr}(g_{i}|u;K).
\label{eq:like}
\end{equation}
The likelihood term \eqref{eq:like} clearly takes different form according to the distribution of the noise. In the following we specify the likelihood associated to the noises considered in this paper.\\

\noindent\textbf{Additive White Gaussian Noise}. If the noise is known to be AWG with standard deviation $\sigma$, the likelihood term in \eqref{eq:like} is as follows:
\begin{equation}
\text{Pr}(g|u;K)=\prod_{i=1}^{n}\,\frac{1}{\sqrt{2\pi}\sigma}\,\text{exp}\bigg(-\frac{(Ku-g)_{i}^{2}}{2\sigma^{2}}\,\bigg)=\frac{1}{W}\,\text{exp}\bigg(-\frac{\lVert Ku-g\rVert_{2}^{2}}{2\sigma^{2}}\,\bigg),
\label{eq:g_like}
\end{equation}
where $W$ is the normalization constant not depending on $u$.
\\Therefore, after replacing our prior \eqref{eq:nsmrf} and the Gaussian likelihood \eqref{eq:g_like} in the MAP inference formula \eqref{eq:map2}, and dropping the constant terms, we obtain our TV$_{p,\alpha}^{\mathrm{sv}}$-L$_{2}$ model in 
\eqref{eq:PMa} with $q=2$, that is in extended form:
\begin{equation}
u^* \:\;{\leftarrow}\;\: \arg \min_{u \in \mathbb{R}^n} \;
\left\{ \, \sum_{i=1}^{n} \alpha_{i} \lVert (\nabla u)_{i} \rVert_{2}^{p_{i}}+\frac{\mu}{2}\lVert Ku-g \rVert_{2}^{2}
\, \right\},
\end{equation}
where we set $\mu=1/\sigma^{2}$.
\\

\noindent \textbf{Additive White Laplace Noise}. If the noise is known to be AWL with scale parameter $\beta$, the likelihood term in \eqref{eq:like} takes the following form:
\begin{equation}
\text{Pr}(g|u;K)=\prod_{i=1}^{n}\,\frac{1}{2\beta}\,\text{exp}\bigg(-\frac{|Ku-g|_{i}}{\beta}\,\bigg)=\frac{1}{W}\,\text{exp}\bigg(-\frac{\lVert Ku-g\rVert_{1}}{\beta}\,\bigg),
\label{eq:l_like}
\end{equation}
\\Therefore, after replacing our prior \eqref{eq:nsmrf} and the Laplace likelihood \eqref{eq:l_like} in the MAP inference formula \eqref{eq:map2}, and dropping the constant terms, we obtain our TV$_{p,\alpha}^{\mathrm{sv}}$-L$_{1}$ model in 
\eqref{eq:PMa} with $q=1$, that is in extended form:
\begin{equation}
	u^* \:\;{\leftarrow}\;\: \arg \min_{u \in \mathbb{R}^n} \;
	\left\{ \, \sum_{i=1}^{n} \alpha_{i} \lVert (\nabla u)_{i} \rVert_{2}^{p_{i}}+\mu\lVert Ku-g \rVert_{1}
	\, \right\},
	\label{eq:modl1}
\end{equation}
where we set $\mu=1/\beta$.\\

\noindent \textbf{Salt and Pepper noise}. The SPN can be classified as a \emph{sparse} noise, since it corrupts only a subset of pixels according to \eqref{eq:SDM} . In this case, in order to strongly promote the sparsity of the noise, a popular choice is to adopt the $\ell_{0}$ pseudo-norm of the residual $Ku-g$ as the fidelity term. Nevertheless, it is very common to substitute the $\ell_{0}$ pseudo-norm with the $\ell_{1}$ norm, which is easier to deal with - since it is convex - and still allows a good sparsification effect. Hence, also in this case, the problem to which we are referring is \eqref{eq:modl1}.

\section{Estimation of the space-variant parameters}
\label{sec:pi}
The proposed regularization term \eqref{eq:PMb} is derived by assuming that, for each pixel position $i=1,...,n$, the magnitude - that is, the $\ell_2$ norm - of the gradients of the target image in a surrounding neighborhood distributes according to a half-Generalized Gaussian (hGG) distribution, whose probability density function is given in \eqref{eq:hgg}. This means that the distribution of the $\ell_{2}$ norm of the gradients in the target image is defined pixel-wise as follows:
\begin{equation}
\text{Pr}\big(\,\lVert (\nabla u)_{i} \rVert\,;p_i,\alpha_i \,\big) \:\;{=}\;\: 
\frac{\alpha_i p_i}{\Gamma(1/p_i)} \, \exp\big(-\big(\,\alpha_i\, \lVert (\nabla u)_{i} \rVert\,\big)^{p_i}\,\big).
\end{equation}

\noindent In order to use the proposed regularization term, we thus need to generate the $p$-map and the $\alpha$-map.\\
The method proposed in \cite{tvpl2} for estimating a global, image-based $p$ value requires a very large number of samples in order to provide statistically reliable estimates, therefore it could not be generalized to our proposal since we use small size image neighborhoods for the estimation of local $p$ values. In \cite{vip} the authors proposed a new method based on the statistical inference procedure illustrated in \cite{shape2} which is sufficiently robust to our purposes. For completeness, in the following we briefly outline the method.

Let $u \in \mathbb{R}^n$ be the vectorized form of an image for which we want to estimate the associated
vector of space-variant parameters $p_i$, $i \in \Omega$.
First, we compute the vector $m \in \mathbb{R}^n$ containing the magnitudes of the gradients of the image $u$; in formulas:
\begin{equation}
m_i \;{:=}\; \left\| (\nabla u)_{i} \right\|_2, \quad\; i \in \Omega \, .
\label{eq:m}
\end{equation}
Then, we estimate each parameter $p_i$ by applying the statistical inference technique in \cite{shape2}
to the local data set consisting of the computed gradient magnitudes in a neighborhood of the
pixel $i$. In particular, we use symmetric square neighborhoods $N_{\,i}^{\,s}$ of size $s \in \{3,5,\ldots\}$ 
centered at pixel $i \in \Omega$.
%
%
Following \cite{shape2}, the values $p_i$, $i \in \Omega$, shape parameters of the hGG distributions, 
are estimated as follows:
\begin{equation}
p_i \,\;{=}\;\, h^{-1}(\rho_i), \quad\;
\rho_i \,\;{=}\;\;
\mathrm{card}\big(N_{\,i}^{\,s}\big) \,
\bigg( \sum_{j \in N_{\,i}^{\,s}} \! m_j^2 \bigg)
\, / \,
\bigg( \sum_{j \in N_{\,i}^{\,s}} \! | m_j | \bigg)^{\!\!2} ,
\quad i \in \Omega \, ,
\label{eq:pi_est}
\end{equation}
where $\mathrm{card}(A)$ denotes the cardinality of set $A$ and where the function $h: ]0,+\infty[ \to ]0,+\infty[$, 
referred to as the \emph{generalized Gaussian ratio function} in \cite{shape2}, is defined by
\begin{equation}
h(z) \,\;{=}\;\, \big( \Gamma(1/z) \,\, \Gamma(3/z) \big) \, / \, \big( \Gamma^2(2/z) \big) \, , 
\label{eq:h}
\end{equation}
with $\Gamma(\,\cdot\,)$ indicating the Gamma function \cite{Gamma}. The function $h$ in (\ref{eq:h}) 
is continuous, monotonically decreasing and surjective, hence invertible.
Moreover, since $h$ is not data-dependent, its inverse $h^{-1}$, representing the values $p_{i}$, can be pre-computed
off-line and stored as a lookup-table, restricted to $(0,2]$, such that at run-time the final
step of the estimation in (\ref{eq:pi_est}) can be carried out very efficiently.

The key novelty of our proposal relies on exploiting all the advantages of using a space-variant hGG distribution model, hence  
we compute also the map of local scale parameters $\alpha_i$, $i = 1,\ldots,n$.
We propose to estimate such scale parameters by means of a Maximum Likelihood approach. Once $p_i$ for a pixel is estimated, the local likelihood function is given by

\begin{eqnarray}
\mathcal{L}(\alpha,p_i;x_{1},...,x_{n})
&\;\;{=}\;\;&
\displaystyle{
	\prod_{i = 1}^{n} \bigg(\frac{\alpha p_i}{\Gamma(1/p_i)}\bigg) \exp(-(\alpha x_{i})^{p_i})} \nonumber \\
&\;\;{=}\;\;&\displaystyle{
	\bigg(\frac{\alpha p_i}{\Gamma(1/p_i)}\bigg)^{n} \exp\bigg(-\sum_{i=1}^{n}(\alpha x_{i})^{p_i}\bigg),
}
%
\label{eq:lklh}
\end{eqnarray}
such that the value of the local scale parameter is obtained by maximizing (\ref{eq:lklh}), that is by solving the following optimization problem:

\begin{eqnarray}
\alpha_i
&\;\;{=}\;\;&
\displaystyle{
	\text{arg}\max_{\alpha} \log \mathcal{L}(\alpha,p_i;x_{1},...,x_{n})}
\nonumber \\
&\;\;{=}\;\;&
\displaystyle{\text{arg} \max_{\alpha} \bigg\{n\log \alpha - \sum_{i=1}^{n} (\alpha	x_{i})^{p_i} \bigg\}.
}
%
\label{eq:minpb}
\end{eqnarray}
By imposing the first order optimality condition for problem \eqref{eq:minpb}, we obtain the following closed form estimation formula:
\begin{equation}
\alpha_i=\bigg(\frac{p_i}{n}\sum_{i=1}^{n}x_{i}^{p_i}\bigg)^{-\frac{1}{p_i}}.
\label{eq:alpha_i}
\end{equation}

\begin{figure}[tbh]
	\center
	\begin{tabular}{ccc}
		\includegraphics[width=1.2in]{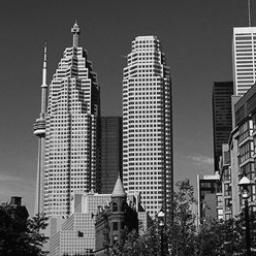} &
		\includegraphics[width=1.2in]{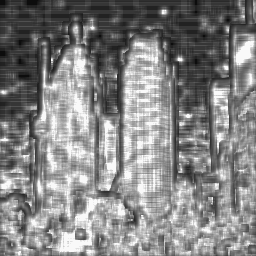} &
		\includegraphics[width=1.2in]{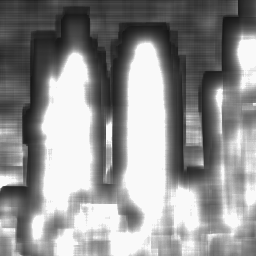}\\
		(a)&(b)&(c)\\
	\end{tabular}	
	
	\caption{Original test image \texttt{skyscraper} (a), $p$-map for $s=3$ (b) and $s=10$ (c).}
	\label{fig:map}	
\end{figure}

In \mbox{Fig. \ref{fig:map}} the maps of local $p$ values, obtained with neighborhoods of size $s=3$ (b) and $s=11$ (c) starting from the original test image \texttt{skyscraper} (a) are shown. Both maps are scaled in the same range for visual comparison. As the size $s$ increases, image features of increasing scale are highlighted, but in any case the method associates very low $p$ values with flat regions and higher 
values with edges. It is worth remarking that in Sect. \ref{sec:nr} numerical experiments have been carried out by computing the $p$-map starting from the corrupted images.

\begin{figure}[tbh]
	\center
	\begin{tabular}{ccc}
		\includegraphics[width=1.2in]{sky_true.png} &
		\includegraphics[width=1.2in]{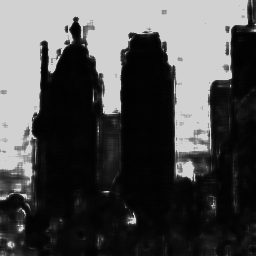} &
		\includegraphics[width=1.2in]{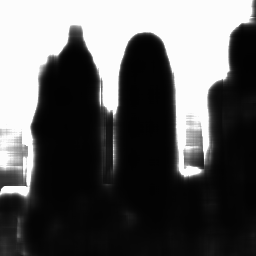}\\
		(a)&(b)&(c)\\
	\end{tabular}	
	
	\caption{Original test image \texttt{skyscraper} (a), $\alpha$-map for $s=3$ (b) and $s=10$ (c).}
	\label{fig:mapscale}	
\end{figure}

In \mbox{Fig. \ref{fig:mapscale}} the maps of local $\alpha$ values, obtained with neighborhoods of size $s=3$ (b) and $s=11$ (c) starting from the original test image \texttt{skyscraper} (a) are shown. Both maps are scaled in the range $[0,1]$ for visual comparison.

\section{Numerical solution by ADMM}
\label{sec:admm}
In this section, we illustrate the ADMM-based iterative algorithm used to numerically
solve the proposed model (\ref{eq:PMa})--(\ref{eq:PMb}) for both cases $q = 2$ and $q = 1$.
To this purpose, first we resort to the variable splitting technique \cite{VAR_SPL1}
and introduce two auxiliary variables $r \in \mathbb{R}^n$ and $t \in \mathbb{R}^{2n}$, 
such that model (\ref{eq:PMa})--(\ref{eq:PMb}) is rewritten
in the following equivalent constrained form:
\begin{eqnarray}
\{ \, u^*,r^*,t^* \}
\:\;{\leftarrow}\;\:
\mathrm{arg}
\min_{u,r,t}
&\:&\bigg\{ \:
\sum_{i = 1}^{n} \alpha_i \| t_i \|_2^{p_i}
\;{+}\;
\frac{\mu}{q} \, \| r \|_q^q
\: \bigg\} ,
\quad  q \in \{1,2\} \, ,
\label{eq:PM_ADMM_a} \vspace {0.2cm} \\
\mathrm{subject}\:\mathrm{to:}
&& \; r \;{=}\; K u - g \, , \;\: t \;{=}\; D u \, ,
\label{eq:PM_ADMM_b}
\end{eqnarray}
where $D := (D_h^T,D_v^T)^T \in \mathbb{R}^{2n \times n}$ and $t_i \:{:=}\: \big( (D_h u)_i \,,\, (D_v u)_i \big)^T \in \mathbb{R}^2$
represents the discrete gradient of image $u$ at pixel $i$.
%
%
%

To solve problem (\ref{eq:PM_ADMM_a})--(\ref{eq:PM_ADMM_b}) by ADMM \cite{BOYD_ADMM}, we define the augmented Lagrangian functional
\begin{eqnarray}
\mathcal{L}(u,r,t;\lambda_r,\lambda_t)
&\;\;{=}\;\;&
\displaystyle{
	\sum_{i = 1}^{n} \alpha_i \| t_i \|_2^{p_i}
	\;{+}\;
	\frac{\mu}{q} \, \| r \|_q^q
	\,{-}\; \langle \, \lambda_t , t - D u \, \rangle
	\;{+}\;
	\frac{\beta_t}{2} \: \| t - D u \|_2^2
} \nonumber \\
&&\displaystyle{
	{-}\; \langle \, \lambda_r , r - (Ku-g) \, \rangle
	\,\;\;{+}\;
	\frac{\beta_r}{2} \, \| \, r - (Ku-g) \|_2^2 \, ,
}
%
\label{eq:PM_AL}
\end{eqnarray}
where $\beta_r, \beta_t > 0$ are scalar penalty parameters and $\lambda_r \in \mathbb{R}^n$, $\lambda_t \in \mathbb{R}^{2n}$
are the vectors of Lagrange multipliers associated with the linear constraints $r = Ku-g$ and $t = Du$
in (\ref{eq:PM_ADMM_b}), respectively.
%
%

Solving (\ref{eq:PM_ADMM_a})--(\ref{eq:PM_ADMM_b}) is thus equivalent to seek for the solutions of the following saddle point problem:
\begin{eqnarray}
\mathrm{Find}&&
\;\, (x^*;\lambda^*)
\;\;{\in}\;\;
X \times \Lambda \nonumber \\
\mathrm{such}\;\mathrm{that}&&
\; \mathcal{L}(x^*;\lambda)
\:\;{\leq}\;\;
\mathcal{L}(x^*;\lambda^*)
\:\;{\leq}\;\;
\mathcal{L}(x;\lambda^*) 
\;\;\;\;\: \forall \: (x;\lambda)
\;\;{\in}\;\;
X \times \Lambda
\: ,
\label{eq:PM_new_S}
\end{eqnarray}
with the augmented lagrangian functional $\mathcal{L}$ defined in (\ref{eq:PM_AL}) and where, for simplicity of notations, we set $x := (u,r,t)$, 
$\lambda := (\lambda_r,\lambda_t)$, 
$X := \mathbb{R}^n \times \mathbb{R}^n \times \mathbb{R}^{2n}$ and $\Lambda := \mathbb{R}^n \times \mathbb{R}^{2n}$.

Given the previously computed (or initialized for $k = 0$) vectors $u^{(k)}$, $\lambda_r^{(k)}$
and $\lambda_t^{(k)}$, the $k$-th iteration of the proposed ADMM-based iterative scheme applied to the solution
of the saddle-point problem (\ref{eq:PM_new_S}) - minimization for the primal
variables $u,r,t$, maximization for the dual variables $\lambda_r,\lambda_t$ - reads as follows:
\begin{eqnarray}
&
r^{(k+1)} &
\;{\leftarrow}\;\;\,\,
\mathrm{arg} \: \min_{r \in \mathbb{R}^n} \;
\mathcal{L}(u^{(k)},r,t^{(k)};\lambda_r^{(k)},\lambda_t^{(k)}) \, ,
\label{eq:PM_ADMM_r} \\
&
t^{(k+1)} &
\;{\leftarrow}\;\;\,\,
\mathrm{arg} \: \min_{t \in \mathbb{R}^{2n}} \;
\mathcal{L}(u^{(k)},r^{(k+1)},t;\lambda_r^{(k)},\lambda_t^{(k)}) \, ,
\label{eq:PM_ADMM_t} \\
&
u^{(k+1)} &
\;{\leftarrow}\;\;\,\,
\mathrm{arg} \: \min_{u \in \mathbb{R}^n} \;
\mathcal{L}(u,r^{(k+1)},t^{(k+1)};\lambda_r^{(k)},\lambda_t^{(k)}) \, ,
\label{eq:PM_ADMM_u} \\
&
\lambda_r^{(k+1)} &
\;{\leftarrow}\;\;\,\,
\lambda_r^{(k)} \;{-}\; \beta_r \, \big( \, r^{(k+1)} \;{-}\; (K u^{(k+1)}-g) \, \big) \, ,
\label{eq:PM_ADMM_lr} \\
&
\lambda_t^{(k+1)} &
\;{\leftarrow}\;\;\,\,
\lambda_t^{(k)} \;{-}\; \beta_t \, \big( \, t^{(k+1)} \;{-}\; D u^{(k+1)} \, \big) \, .
\label{eq:PM_ADMM_lt}
\end{eqnarray}
In the following three subsections we describe how to solve the minimization sub-problems (\ref{eq:PM_ADMM_r}), 
(\ref{eq:PM_ADMM_t}) and (\ref{eq:PM_ADMM_u}) for the primal variables $r$, $t$ and $u$, respectively, 
in both cases $q \in \{1,2\}$.
In particular, we remark that thanks to the preliminary ADMM variable splitting procedure, sub-problems
(\ref{eq:PM_ADMM_t}) and (\ref{eq:PM_ADMM_u}) for the variables $t$ and $u$ are identical in the two cases $q \in \{1,2\}$ and their solution can be obtained based on formulas given in~\cite{tvpl2} for the same sub-problems. 

\subsection{Minimization sub-problem for the primal variable $r$}
\label{subsec:sub_r}
Recalling the definition of the augmented Lagrangian functional in (\ref{eq:PM_AL})
and carrying out some simple algebraic manipulations, the minimization sub-problem (\ref{eq:PM_ADMM_r})
for the primal variable $r$ can be written as 
\begin{eqnarray}
r^{(k+1)}
&\;{\leftarrow}\;&
\mathrm{arg} \min_{r \in \mathbb{R}^n}\:
\left\{ \,
\frac{\mu}{q} \, \| r \|_q^q
\;{+}\;
\frac{\beta_r}{2} \,
\big\| r - v^{(k)} \big\|_2^2
\: \right\} \, , 
\quad  q \in \{1,2\} \, ,
\label{eq:sub_r_q12}
\end{eqnarray}
%
%
%
%
with the constant (with respect to the optimization variable $r$) vector $v^{(k)} \in \mathbb{R}^n$ given by
\begin{equation}
v^{(k)} \;{=}\;\: Ku^{(k)} -\, g \: + \, \frac{1}{\beta_r} \, \lambda_r^{(k)} \; .
\label{eq:v_def}
\end{equation}
Since $\mu \geq 0$, $\beta_r>0$, in both cases $q \in \{1,2\}$ the cost function in (\ref{eq:sub_r_q12}) is strongly convex, 
hence it admits a unique global minimizer. 
In particular, the unique solution $r^{(k+1)}$ of (\ref{eq:sub_r_q12}) can be computed, depending on $q$,
by means of the following closed-form formulas:
\begin{eqnarray}
\mathrm{case}\;\;q \;{=}\; 1 \, : \qquad\!
r^{(k+1)}
&\;{=}\;&
\mathrm{sign}\big( v^{(k)} \big) \, \odot \,\,
\max\big\{ \, |v^{(k)}| - \mu / \beta_r \, , \, 0 \, \big\} \: ,
\label{eq:sub_r_q1_sol} \\
\mathrm{case}\;\;q \;{=}\; 2 \, : \qquad\!
r^{(k+1)}
&\;{=}\;&
\big(\beta_r / (\beta_r+\mu)\big) \, v^{(k)} \: ,
\label{eq:sub_r_q2_sol}
\end{eqnarray}
where $\mathrm{sign}(\,\cdot\,)$ and $| \, \cdot \, |$ in (\ref{eq:sub_r_q1_sol}) denote the component-wise
signum and absolute value functions and $\,\odot$ indicates the component-wise vectors product.
We remark that formula (\ref{eq:sub_r_q1_sol}) represents a well-known component-wise
soft-thresholding operator - see e.g. \cite{tvl1} - whereas (\ref{eq:sub_r_q2_sol}) comes easily
from first-order optimality conditions of (\ref{eq:sub_r_q12}).

In case that the regularization parameter $\mu$ is regarded as a constant - that is, it is fixed a priori -
then formulas (\ref{eq:sub_r_q1_sol})--(\ref{eq:sub_r_q2_sol}) allow to determine very efficiently
the solution $r^{(k+1)}$ of this sub-problem.
However, as previously stated, in the case $q = 2$ we aim also at
automatically adjusting $\mu$ along iterations - that is, $\mu$ becomes $\mu^{(k)}$ - such that the final solution $u^*$ of our model
(\ref{eq:PMa})--(\ref{eq:PMb}) satisfies the discrepancy principle \cite{WC12}.
To this aim, in the following we propose a procedure which builds upon those presented in \cite{APE,JMIV16} but, due to a different ADMM initial variable splitting, needs to be adapted and is worth to be outlined in detail.
%
%

We consider the discrepancy associated with the solution $r^{(k+1)}$ in (\ref{eq:sub_r_q2_sol}) as
a function $\delta^{(k+1)}: [0,+\infty[ \rightarrow [0,+\infty[$ of the regularization parameter $\mu$:
\begin{equation}
\delta^{(k+1)}(\mu)
\,\;{:=}\;\,
\big\| r^{(k+1)}(\mu) \big\|_2
\;{=}\;\:
\frac{\beta_r}{\beta_r+\mu} \, \big\| \, v^{(k)} \big\|_2 \; ,
\label{eq:d_und}
\end{equation}
where the second equality comes from (\ref{eq:sub_r_q2_sol}).
The discrepancy function in
(\ref{eq:d_und}) is clearly continuous, non-negative and monotonically decreasing over its
entire domain \mbox{$\mu \in [0,+\infty[$} and at the extremes we have
$\,\delta^{(k+1)}(\mu=0) = \| v^{(k)} \|_2$,
$\,\delta^{(k+1)}(\mu \to +\infty) = 0$.
In order to set a value $\mu^{(k+1)}$
such that the discrepancy principle is satisfied here for the auxiliary variable
$r$ (recall that $r=Ku-g$ represents the residue of the restoration), we consider two complementary
cases based on the value of the norm of the vector $v^{(k)}$ defined in (\ref{eq:v_def}).

In case that $\,\| \, v^{(k)} \|_2 \leq \bar{\delta}$, with $\bar{\delta}$ denoting the noise level,
then from (\ref{eq:d_und}) and from the fact that $\,0 < \beta_r / (\beta_r + \mu) \leq 1$, it 
follows that $\,\delta^{(k+1)}(\mu) \;{\leq}\; \bar{\delta} \;\: \forall \, \mu \in [0,+\infty[$, that is the
discrepancy principle is satisfied for any $\mu$.
In this case we thus set $\mu^{(k+1)} = 0$, such that, according to (\ref{eq:sub_r_q2_sol}),
the sub-problem solution is $r^{(k+1)} = v^{(k)}$.

In case that $\,\| v^{(k)} \|_2 > \bar{\delta}$, the properties of the discrepancy
function $\delta^{(k+1)}(\mu)$ in (\ref{eq:d_und}) guarantee that there exists a unique value $\mu^{(k+1)}$ of $\,\mu$ such that $\delta^{(k+1)}(\mu^{(k+1)}) = \bar{\delta}$.
Recalling (\ref{eq:d_und}), we have
$\big(\beta_r / (\beta_r+\mu^{(k+1)})\big) \| \, v^{(k)} \|_2 \;{=}\; \bar{\delta}
\:\;\;\;{\Longleftrightarrow}\;\;\;
\mu^{(k+1)} = \beta_r \big(\, \| \, v^{(k)} \|_2 / \bar{\delta} \:\;{-}\; 1 \,\big)$.
Replacing this expression 
for $\mu$ in (\ref{eq:sub_r_q2_sol}), 
the sub-problem solution is 
$\,r^{(k+1)} \;{=}\;\: \bar{\delta} \, v^{(k)} / \| \, v^{(k)} \|_2$.

To summarize, the solution of this sub-problem at any iteration $k$ is computed by (\ref{eq:sub_r_q1_sol}) 
for the case $q = 1$ whereas for the case $q = 2$ it is determined as follows:
\begin{equation}
\begin{array}{llll}
\| \, v^{(k)} \|_2 \;{\leq}\; \bar{\delta} &
\;\Longrightarrow\; &
\mu^{(k+1)} \;{=}\; 0 , & 
\;\; r^{(k+1)} \;{=}\; v^{(k)} \vspace{0.2cm} \\
%
\| \, v^{(k)} \|_2 \;{>}\; \bar{\delta} &
\;\Longrightarrow\; &
\mu^{(k+1)} \;{=}\; \beta_r \big( \| \, v^{(k)} \|_2 / \bar{\delta} - 1 \big) , &
\;\; r^{(k+1)} \;{=}\; \bar{\delta} \,\, v^{(k)} / \| \, v^{(k)} \|_2
%
\end{array}
\label{eq:sub_r_q2_q2_sol}
\end{equation}

\subsection{Minimization sub-problem for the primal variable $t$}
\label{subsec:sub_t}
Given the definition of the augmented Lagrangian functional in (\ref{eq:PM_AL}), the minimization sub-problem for 
the primal variable $t$ in (\ref{eq:PM_ADMM_t}) can be written as follows:
\begin{eqnarray}
&t^{(k+1)}& \;{\leftarrow}\; \mathrm{arg} \, \min_{t \in \mathbb{R}^{2n}}\:
\left\{ \:
\sum_{i=1}^{n} \alpha_i \left\| t_i \right\|_2^{p_i}
\;{-}\;
\langle \lambda_t^{(k)} , t - D u^{(k)} \rangle
\;{+}\;
\frac{\beta_t}{2}
\left\| t - D u^{(k)} \right\|_2^2
\:\right\} \nonumber \\
&& \;{\leftarrow}\; \mathrm{arg} \, \min_{t \in \mathbb{R}^{2n}}\:
\left\{ \:
\sum_{i=1}^{n} \alpha_i \left\| t_i \right\|_2^{p_i}
\;{+}\;
\frac{\beta_t}{2}
\left\| t - \left( D u^{(k)} + \frac{1}{\beta_t} \lambda_t^{(k)} \right) \right\|_2^2
\:\right\} \nonumber \\
&& \;{\leftarrow}\; \mathrm{arg} \, \min_{t \in \mathbb{R}^{2n}}\:
\;
\sum_{i=1}^{n}
\left\{
\alpha_i \left\| t_i \right\|_2^{p_i}
\;{+}\;
\frac{\beta_t}{2}
\left\| t_i - \left( \left( D u^{(k)} \right)_i + \frac{1}{\beta_t} \left( \lambda_t^{(k)} \right)_i \right) \right\|_2^2
\right\} \label{eq:sub_t_c}
\: .
\end{eqnarray}
Note that in (\ref{eq:sub_t_c}) the minimized functional is written in explicit component-wise (or pixel-wise) form, with $\left( D u^{(k)} \right)_i, \left( \lambda_t^{(k)} \right)_i \in \mathbb{R}^2$ denoting the discrete gradient and the Lagrange multipliers at pixel $i$, respectively.
Solving the $2n$-dimensional minimization problem in (\ref{eq:sub_t_c}) is thus equivalent to solve the $n$ following independent $2$-dimensional problems:
\begin{eqnarray}
&t^{(k+1)}_i& {\leftarrow}\; \mathrm{arg} \min_{t_i \in \mathbb{R}^2}
\left\{ \,
\left\| t_i \right\|_2^{p_i}
\;{+}\;
\frac{(\beta_t/\alpha_i)}{2} \left\| t_i - q_i^{(k)} \right\|_2^2
\,\right\}
,
\quad i = 1, \ldots , n \: , \label{eq:sub_t_i}
\end{eqnarray}
with the constant vectors $q^{(k)}_i \in \mathbb{R}^2$ defined by
\begin{equation}
q^{(k)}_i \:\;{:=}\;\: \left( D u^{(k)} \right)_i + \frac{1}{\beta_t} \left( \lambda^{(k)}_t \right)_i \;\: ,
\quad i = 1, \ldots , n \: .
\label{eq:sub_t_i_q}
\end{equation}

The solutions of the $n$ optimization problems in (\ref{eq:sub_t_i}) can be obtained based on the results reported in Proposition 1 of \cite{tvpl2}, that is:
\begin{equation}
t_i^{(k+1)}
\;{=}\;\:
\xi_i^{(k)} \, q_i^{(k)} \: ,
\quad i = 1, \ldots , n \: ,
\label{eq:sub_t_i_sol_b}
\end{equation}
where, in particular, the shrinkage coefficients $\,\xi_i^{(k)} \in [0 , 1], \;\, i = 1,\ldots,n ,\,$ are given by formulas (50)--(52) 
in \cite{tvpl2}.

The overall computational cost of this subproblem is linear in the number of pixels $n$.

\subsection{Minimization sub-problem for the primal variable $u$}
\label{subsec:sub_u}
As illustrated in~\cite{tvpl2}, the minimization sub-problem (\ref{eq:PM_ADMM_u}) for the primal variable $u$ reduces to 
the solution of the following $n \times n$ system of linear equations
\begin{equation}
\left(
D^T D
+ \frac{\beta_r}{\beta_t} K^T K
\right)
u
=
D^T \left( t^{(k+1)} - \frac{1}{\beta_t} \lambda^{(k)}_t \right)
+
\frac{\beta_r}{\beta_t} K^T \left( r^{(k+1)} - \frac{1}{\beta_r} \lambda^{(k)}_r + g  \right)
\: ,
\label{eq:sub_u_sol}
\end{equation}
%
%
which is solvable if the coefficient matrix has full-rank, that is if the following condition holds:
\begin{equation}
\mathrm{Ker}\left( D^T D \right) 
\;{\cap}\;
\mathrm{Ker}\left( K^T K \right) 
\;{=}\;
\{ 0 \}
\: ,
\label{eq:sub_u_solv}
\end{equation}
where $\mathrm{Ker}(M)$ denotes the null space of matrix $M$ and $0$ is the $n$-dimensional null vector. 
In our case, condition (\ref{eq:sub_u_solv}) is satisfied. In fact, $K$ represents a blurring operator,
which is a low-pass filter, whereas the regularization matrix $D$ is a first-order difference
operator and, hence, is a high-pass filter. Moreover, since $\beta_t, \beta_r > 0$, the coefficient matrix in (\ref{eq:sub_u_sol}) 
is symmetric positive definite and typically highly sparse. Hence, the linear system in (\ref{eq:sub_u_sol}) can be solved quite efficiently by the iterative 
(eventually preconditioned) conjugate gradient method. 
Moreover, under appropriate assumptions about the solution $u$ near the image boundary, the linear system can be solved even more efficiently. 
We assume \emph{periodic} boundary conditions for $u$, so that both $D^T D$ and $K^T K$ are block circulant matrices with circulant blocks and, hence, the coefficient matrix in (\ref{eq:sub_u_sol}) can be diagonalized by the 2D discrete Fourier transform (FFT implementation). 
Provided that the penalty parameters $\beta_t$, $\beta_r$ are kept fixed during the ADMM iterations, the coefficient matrix in (\ref{eq:sub_u_sol}) does not change and it can be diagonalized once for all at the beginning. Therefore, at any ADMM iteration the linear system (\ref{eq:sub_u_sol}) can be solved by one forward 2D FFT and one inverse 2D FFT, each at a cost of $O(n \log n)$.


%
%
%
\subsection{ADMM-based iterative scheme}
\label{subsec:alg}

To summarize previous results, in Algorithm \ref{alg:1} we report the main steps of the proposed ADMM-based iterative
scheme used to solve the saddle-point problem (\ref{eq:PM_AL})--(\ref{eq:PM_new_S}) and, hence, to compute solutions of 
the proposed model (\ref{eq:PMa})--(\ref{eq:PMb}).


In the field of image and signal processing the ADMM has been one of the most powerful and successful methods for solving various convex or nonconvex optimization problems.
In convex settings, numerous convergence results have been established for ADMM as well as its varieties, see for example \cite{HY} and references therein. 
In particular, convergence results cover the proposed TV$_{p,\alpha}$-L$_q$ models, $q \in \{1,2\}$, in the special case of $ p_i \ge 1 \; \forall \, i$.
However, in case that one or more $p_i < 1$, the ADMM is under nonconvex settings, where there have been a few studies on the convergence properties. 
To the best of our knowledge, existing convergence results of ADMM for nonconvex problems is very limited
to particular classes of problems and under certain conditions of the dual step size \cite{HLR}.
Nevertheless, the ADMM works extremely well for various applications involving nonconvex optimization problems, and this is a practical 
justification of its wide use.

%
\begin{algorithm}
\caption{ADMM-based scheme for models (\ref{eq:PMa})--(\ref{eq:PMb}) \vspace{0.05cm} }
\label{alg:1}
\vspace{0.2cm}
{\renewcommand{\arraystretch}{1.2}
\renewcommand{\tabcolsep}{0.0cm}
\vspace{-0.08cm}
\begin{tabular}{ll}
\textbf{input}:      & observed image $\,g \:{\in}\; \mathbb{R}^n$ \vspace{0.04cm} \\
\textbf{output}: $\;\:$    & approximate solution $\,u^* {\in}\; \mathbb{R}^n$ of (\ref{eq:PMa})--(\ref{eq:PMb}) \vspace{0.2cm} \\
\end{tabular}
}
\hspace{4cm}
\vspace{0.1cm}
{\renewcommand{\arraystretch}{1.2}
\renewcommand{\tabcolsep}{0.0cm}

\begin{tabular}{rcll}
1. & $\quad$ & \multicolumn{2}{l}{\textbf{initialize:}} \vspace{0.05cm}\\
2. && \multicolumn{2}{l}{$\;\;\;\;\cdot$ estimate parameters $\:p_i$ and $\alpha_i$, $\,i = 1,\ldots,n$, by (\ref{eq:pi_est}) and (\ref{eq:alpha_i}), respectively} \vspace{0.1cm}\\
3. && \multicolumn{2}{l}{$\;\;\;\;\cdot$ set $\:u^{(0)} = g$, $\lambda_r^{(0)} = \lambda_t^{(0)} = 0$} \vspace{0.2cm}\\
2. && \multicolumn{2}{l}{\textbf{for} $\;$ \textit{k = 1, 2, 3, $\, \ldots \,$ until convergence $\:$} \textbf{do}:} \vspace{0.1cm}\\
3. && $\quad\;\;\bf{\cdot}$ \textbf{update primal variables:} &  \vspace{0.05cm} \\
4. && $\qquad\qquad\cdot$ compute $\:r^{(k+1)}$   & by (\ref{eq:v_def}) and (\ref{eq:sub_r_q1_sol}) for $q=1$, (\ref{eq:sub_r_q2_q2_sol}) for $q=2$ \vspace{0.05cm} \\
5. && $\qquad\qquad\cdot$ compute $\:t^{(k+1)}$   & by (\ref{eq:sub_t_i_q}), (\ref{eq:sub_t_i_sol_b}) and formulas (50)--(52) in \cite{tvpl2}  \vspace{0.05cm} \\
5. && $\qquad\qquad\cdot$ compute $\:u^{(k+1)}$   & by solving (\ref{eq:sub_u_sol})  \vspace{0.05cm} \\
6. && $\quad\;\;\bf{\cdot}$ \textbf{update dual variables:} &  \vspace{0.05cm} \\
7. && $\qquad\qquad\cdot$ compute $\:\lambda_r^{(k+1)}$, $\:\lambda_t^{(k+1)}$ & by (\ref{eq:PM_ADMM_lr}), (\ref{eq:PM_ADMM_lt}) \vspace{0.1cm} \\
8. && \multicolumn{2}{l}{\textbf{end$\;$for}} \vspace{0.09cm} \\
9. && \multicolumn{2}{l}{$u^* = u^{(k+1)}$}
\end{tabular}
}
\end{algorithm}
\section{Numerical results}
\label{sec:nr}
In this section, we evaluate experimentally the performance of the two proposed models
TV$_{p,\alpha}^{\mathrm{sv}}$-L$_q$, $q \in \{1,2\}$, defined in (\ref{eq:PMa})--({\ref{eq:PMb}), when
applied to the restoration of gray level images synthetically corrupted
by known blur and by AWGN - in the case of TV$_{p,\alpha}^{\mathrm{sv}}$-L$_2$ model - and SPN or AWLN -
in the case of TV$_{p,\alpha}^{\mathrm{sv}}$-L$_1$ model.
In particular, the two proposed models are compared with:
\begin{itemize}
	\item TV-L$_q$, $\;\;\,q \in \{1,2\}$, defined in (\ref{eq:TVLq}); see \cite{ROF}, \cite{ADMTV}; 
	\item TV$_p$-L$_q$, $\;q \in \{1,2\}$, with global $p \in (0,2]$; see \cite{tvpl2}, \cite{vip};
	\item TV$_p^{\mathrm{sv}}$-L$_q$, $q \in \{1,2\}$, with local $p_i \in (0,2]$, $i \in \{1,\ldots,n\}$; see \cite{vip}.
\end{itemize}

For what concerns the preliminary estimation of the $p_i$ and $\alpha_i$ parameters, we directly apply the procedure outlined in Section \ref{sec:pi} to the observed corrupted image $g$ for the AWGN and AWLN cases. Instead, for the SPN case, in order to have a robust estimation of the parameters, a preliminary processing by an adaptive filter is required. In particular, we assume that the position of the pixels corrupted by the SPN is known a priori, otherwise it can be easily detected as suggested in \cite{mila}. We replace the corrupted pixels with the mean of the non-corrupted pixels of its neighborhood. The size of the neighborhood is variable and depends on the percentage $\mathcal{P}$ of non-corrupted pixels in it. If $\mathcal{P}$ is below a fixed threshold $\overline{\mathcal{P}}$ (usually $\overline{\mathcal{P}}=0.4$), then the neighborhood is enlarged, in order to incorporate a greater number of uncorrupted pixels. The obtained image is then used to compute the $p$-map and the $\alpha$-map. The described strategy has been introduced instead of the classic median filter, whose smoothing effects is quite high. Clearly, the same approach is adopted for the TV$_p^{\mathrm{sv}}$-L$_1$ model to estimate the $p$-map only.

The quality of the observed corrupted images $g$ and of the restored images
$u^*$ is measured - in dB - by means of the Blurred Signal-to-Noise Ratio
\[
\;\mathrm{BSNR}(g,u) = 10\log_{10} \frac{\|Ku - E\,[Ku]\|_2^2}{\|g-Ku\|_2^2}
\]
and the Improved Signal-to-Noise Ratio
\[
\mathrm{ISNR}(g,u,u^*) = 10\log_{10}\frac{\|g-u\|_2^2}{\|u^*-u\|_2^2},
\] respectively, with $u$ denoting the original uncorrupted image and $E\,[Ku]$ the average intensity
of the blurred image $Ku$. In general, the larger the ISNR value, the higher the
quality of restoration.

For all the ADMM-based minimization algorithms and for all the tests, the penalty parameters $\beta_t$ and $\beta_r$ are suitably set. 
Moreover, for all tests, the ADMM iterations of all the compared algorithms are
stopped as soon as two successive iterates satisfy
\begin{equation}
\frac{\big\| u^{(k)} - u^{(k-1)} \big\|_{2}}{\big\| u^{(k-1)}\big\|_{2}} \,\;{<}\;\,10^{-4}.
\end{equation}

For the models with the L$_2$ fidelity term, the regularization parameter $\mu$ has been automatically
set based on the discrepancy principle.
For the models with the L$_1$ fidelity term, $\mu$ has been hand-tuned independently in each test so as
to provide the highest possible ISNR value. In the following, we report numerical results concerning the restoration of blurred images corrupted
by AWGN (Example 1) and by impulsive SPN and AWLN (Example 2).
%
%

\smallskip

\textbf{Example 1: restoration of images corrupted by blur and AWGN.} In this example, we evaluate experimentally the performance of the proposed 
TV$_{p,\alpha}^{\mathrm{sv}}$-L$_2$ model on a purely piecewise constant test image - \texttt{geometric} ($256 \times 256$), Fig. \ref{fig:geomand}(a) - and a partially textured test image - \texttt{skyscraper} ($256 \times 256$), Fig. \ref{fig:geomand}(d). Both images have been synthetically corrupted by a Gaussian blur of parameters \verb|band=5| and \verb|sigma=1.0| and by AWGN characterized by different noise levels. 
The $p,\alpha$-maps have been computed by using neighborhoods of size $s=3$. 

In Table \ref{tab:1} the performance of our model are compared in terms of achieved ISNR values with those of the TV-L$_2$, TV$_{p}$-L$_2$ and TV$_p^{\mathrm{sv}}$-L$_2$ models. The good quality of the restored image by our model can be appreciated by a visual inspection of Figs. \ref{fig:geomand}(c),(f) and by comparing the ISNR values reported in Table 1.

\begin{figure}[tbh]
	\center
	\begin{tabular}{ccc}
		\includegraphics[width=1.85in]{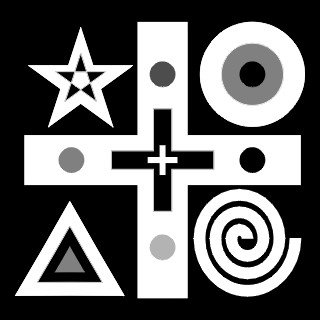} &
		\includegraphics[width=1.85in]{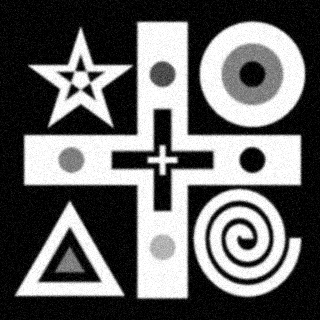} &
		\includegraphics[width=1.85in]{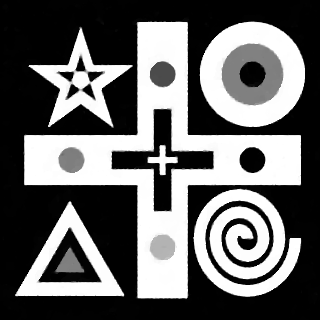}\\
		(a) original &(b) corrupted (BSNR = 20)&(c) TV$_{p,\alpha}^{\mathrm{sv}}$-L$_2$ (ISNR = 8.60)\\
		
		\includegraphics[width=1.85in]{sky_true.png} &
		\includegraphics[width=1.85in]{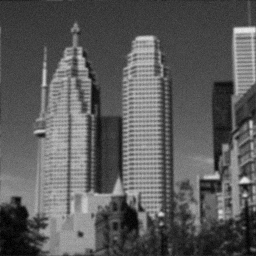} &
		\includegraphics[width=1.85in]{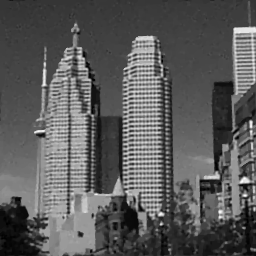}\\
		(d) original &(e) corrupted (BSNR = 20) &(f) TV$_{p,\alpha}^{\mathrm{sv}}$-L$_2$ (ISNR = 3.31)\\
	\end{tabular}
	\caption{Example 1: restoration of the test images \texttt{geometric} and \texttt{skyscraper} corrupted by blur and AWGN.}
	\label{fig:geomand}
\end{figure}

\begin{table}
	\caption{Example 1: achieved ISNR values.}
	\label{tab:1}       
	%
	%
	\begin{tabular}{rrrrr|rrrr}
		\smallskip
		& &&\bf{geometric}  &  & &&\bf{skyscraper}& \\
		\hline\noalign{\smallskip}
		BSNR & TV-L$_2$& TV$_{p}$-L$_2$ & TV$_p^{\mathrm{sv}}$-L$_2$ &TV$_{p,\alpha}^{\mathrm{sv}}$-L$_2$ &TV-L$_2$&  TV$_{p}$-L$_2$ & TV$_p^{\mathrm{sv}}$-L$_2$&  TV$_{p,\alpha}^{\mathrm{sv}}$-L$_2$ \\
		\noalign{\smallskip}\hline\noalign{\smallskip}
		20               & 7.77 & 7.92 & 8.36 & 8.60 & 2.76 &3.00&3.07&3.31\\
		30               & 9.01  & 9.87  & 10.30 & 10.57 & 5.12 &5.52&5.94&6.40 \\
		\noalign{\smallskip}\hline\noalign{\smallskip}
	\end{tabular}
	
\end{table}

\smallskip
\textbf{Example 2: restoration of images corrupted by blur and SPN or AWLN.} In this example we evaluate the performance of the proposed TV$_{p,\alpha}^{\mathrm{sv}}$-L$_1$ model on three medical test images \texttt{lungs} ($468 \times 591$), Fig. \ref{fig:lungs} (a), \texttt{ecography} ($401 \times 511$), Fig. \ref{fig:eco} (a), and \texttt{aneurism} ($701 \times 766$), Fig. \ref{fig:aneu} (a), synthetically corrupted by Gaussian blur of parameters \verb|band=5| and \verb|sigma=1| and by two types of impulsive noise, namely SPN and AWLN. 
The images are provided in the repository at https://medpix.nlm.nih.gov.

\begin{figure}[tbh]
	\center
	\begin{tabular}{ccc}

		\includegraphics[width=2in]{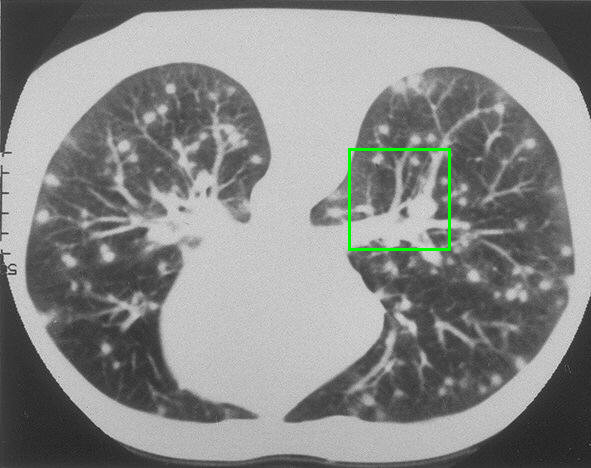} &
		\includegraphics[width=2in]{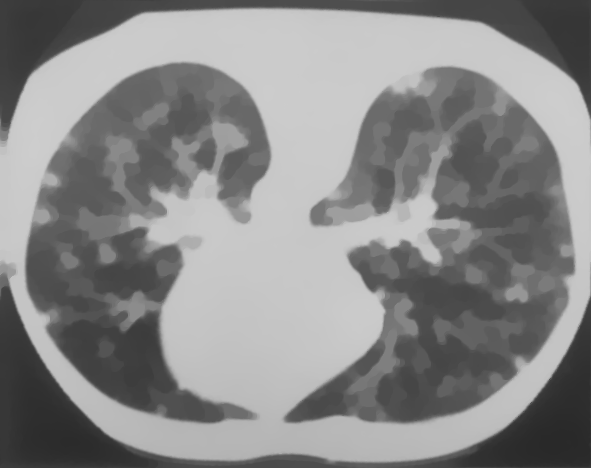} &
		\includegraphics[width=1.6in]{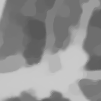}\\
		(a) original&(b) TV-L$_1$ (ISNR = 11.04)&(c) zoom of (b)\\
		\includegraphics[width=2in]{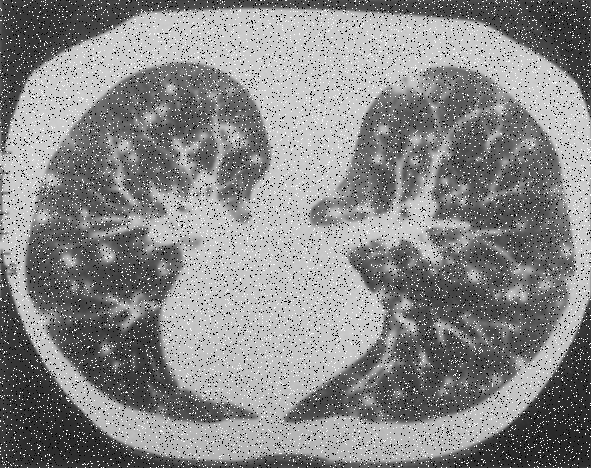} &
		\includegraphics[width=2in]{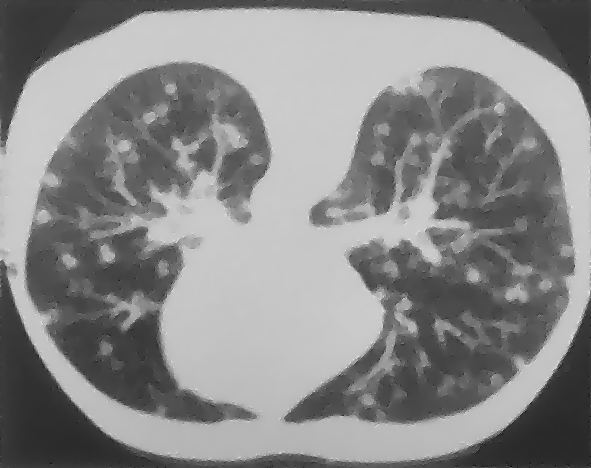} & 
		\includegraphics[width=1.6in]{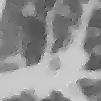}\\
		(d) corrupted&(e) TV$_{p}$-L$_1$ (ISNR = 12.48)&(f) zoom of (e)\\ 
		\includegraphics[width=2in]{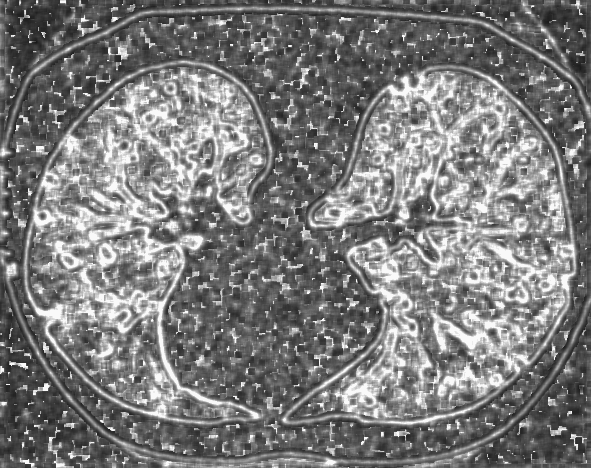} &
		\includegraphics[width=2in]{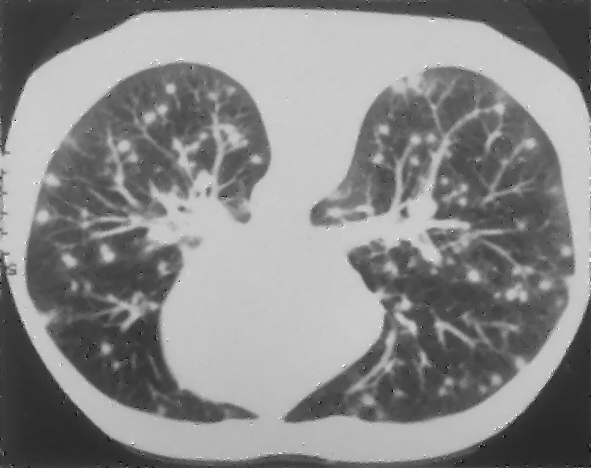} & 
		\includegraphics[width=1.6in]{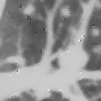}\\
		(g) $p$-map ($s=3$)&(h) TV$_{p}^{\mathrm{sv}}$-L$_1$ (ISNR = 15.30)&(i) zoom of (h)\\ 
		
		\includegraphics[width=2in]{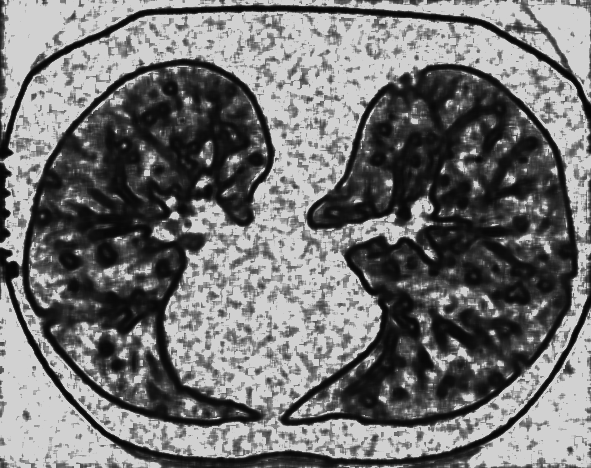} &
		\includegraphics[width=2in]{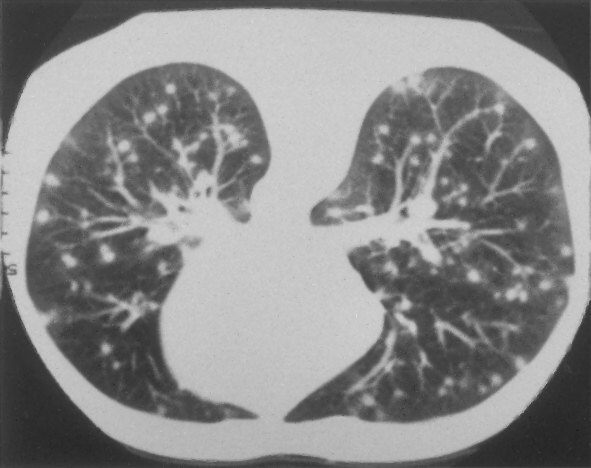} & 
		\includegraphics[width=1.6in]{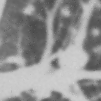}\\
		(l) $\alpha$-map ($s=3$)&(m) TV$_{p,\alpha}^{\mathrm{sv}}$-L$_1$ (ISNR = 16.56)&(n) zoom of (m)\\ 
	\end{tabular}		
	\caption{Example 2 (SPN): visual restoration results for the test image \texttt{lungs} corrupted by a $\gamma=0.1$ level noise.}
	\label{fig:lungs}
\end{figure}

\begin{figure}[tbh]
	\center
	\begin{tabular}{ccc}
		\includegraphics[width=1.95in]{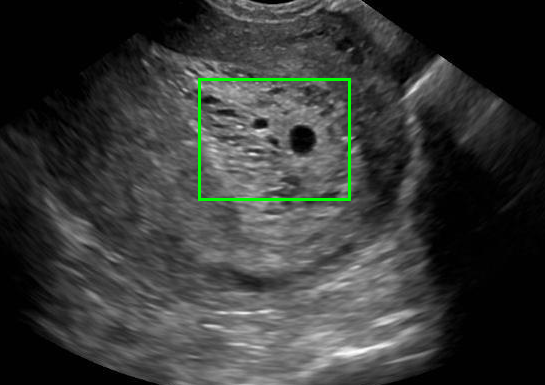} &
		\includegraphics[width=2in]{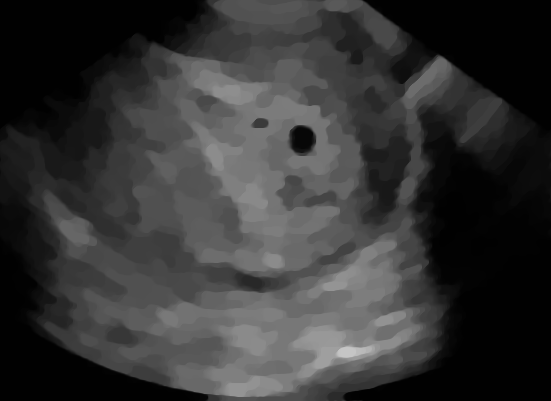} &
		\includegraphics[width=1.75in]{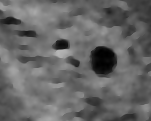}\\
	(a) original&(b) TV-L$_1$ (ISNR = 22.13)&(c) zoom of (b)\\
		\includegraphics[width=1.95in]{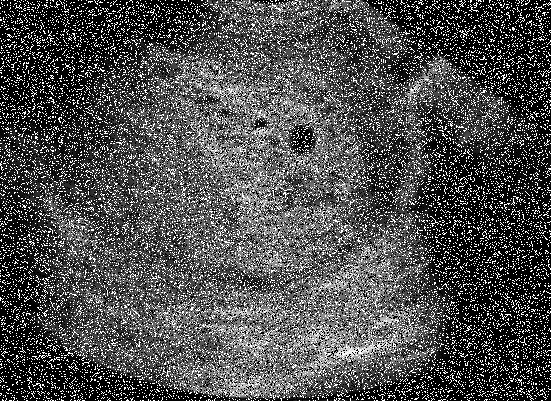} &
		\includegraphics[width=2in]{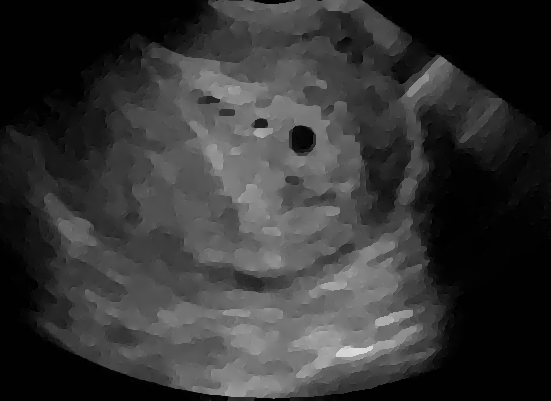} & 
		\includegraphics[width=1.75in]{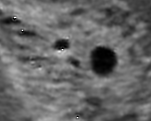}\\
(d) corrupted&(e) TV$_{p}$-L$_1$ (ISNR = 23.15)&(f) zoom of (e)\\ 

		\includegraphics[width=1.95in]{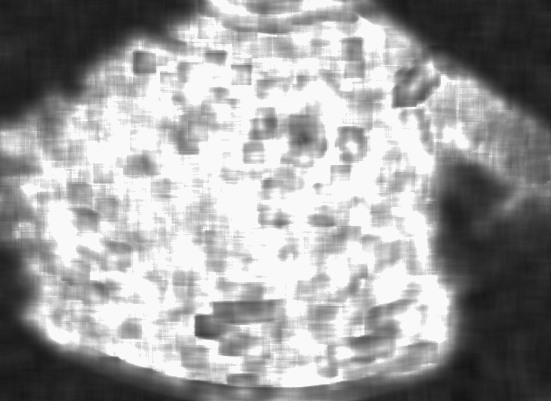} &
		\includegraphics[width=2in]{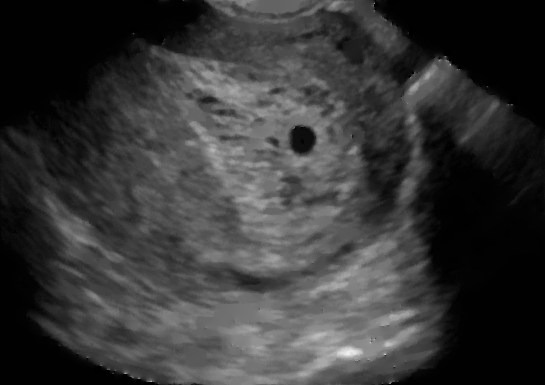} & 
		\includegraphics[width=1.75in]{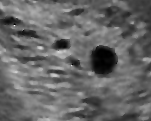}\\
	(g) $p$-map ($s=10$)&(h) TV$_{p}^{\mathrm{sv}}$-L$_1$ (ISNR = 25.46)&(i) zoom of (h)\\

		\includegraphics[width=1.95in]{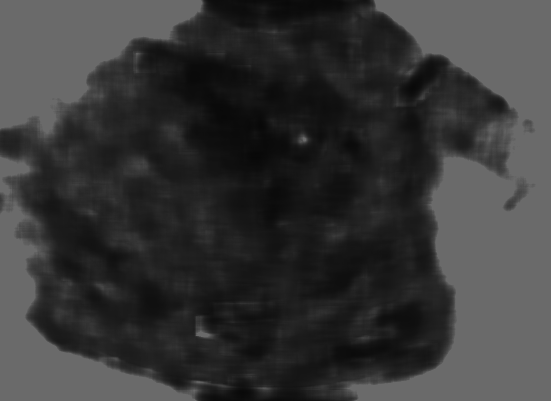} &
		\includegraphics[width=2in]{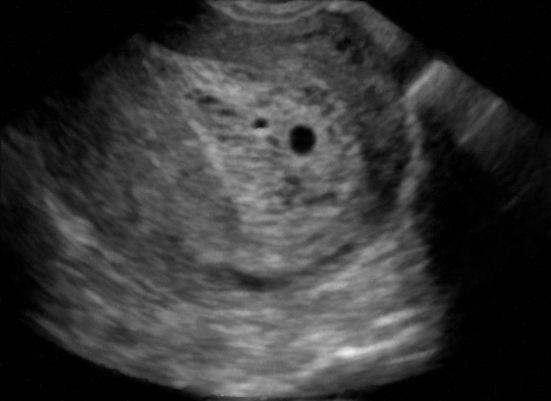} & 
		\includegraphics[width=1.75in]{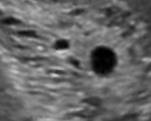}\\
(l) $\alpha$-map ($s=10$)&(m) TV$_{p,\alpha}^{\mathrm{sv}}$-L$_1$ (ISNR = 28.01)&(n) zoom of (m)\\ 

	\end{tabular}		
	\caption{Example 2 (SPN): visual restoration results for the test image \texttt{ecography} corrupted by a $\gamma=0.35$ level noise.}
	\label{fig:eco}
\end{figure}

\begin{figure}[tbh]
	\center
	\begin{tabular}{ccc}
		\includegraphics[width=2in]{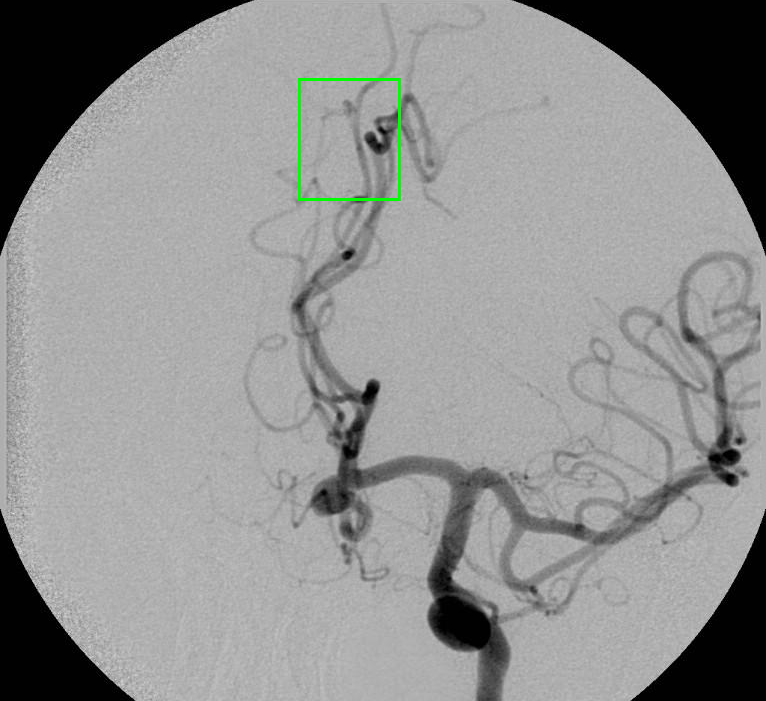} &
		\includegraphics[width=2in]{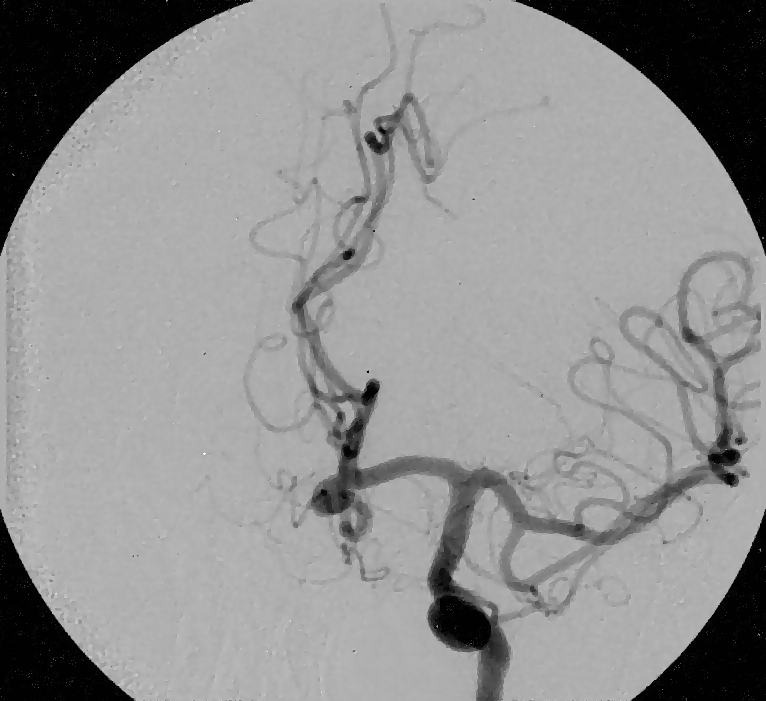} &
		\includegraphics[width=1.53in]{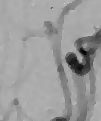}\\
	(a) original&(b) TV-L$_1$ (ISNR = 18.55)&(c) zoom of (b)\\
		\includegraphics[width=2in]{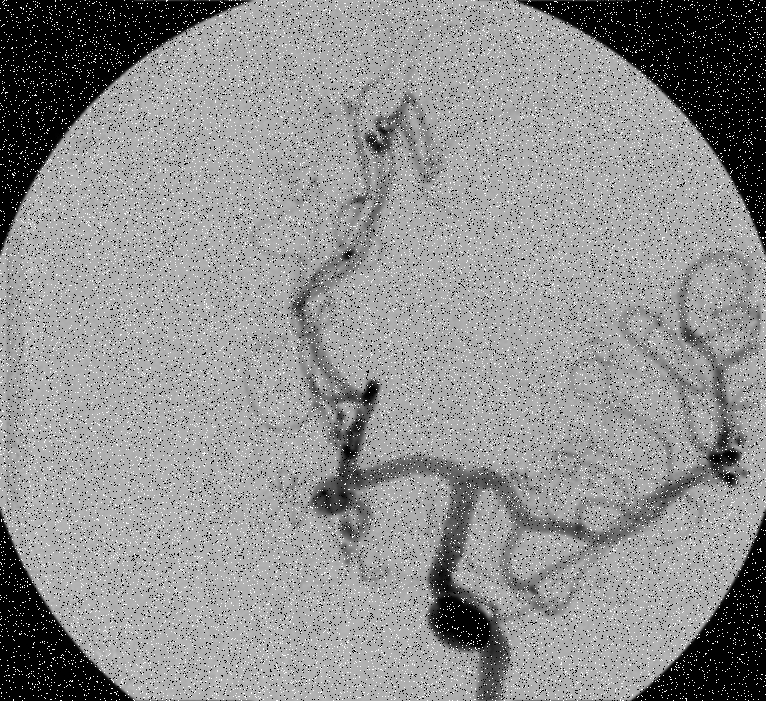} &
		\includegraphics[width=2in]{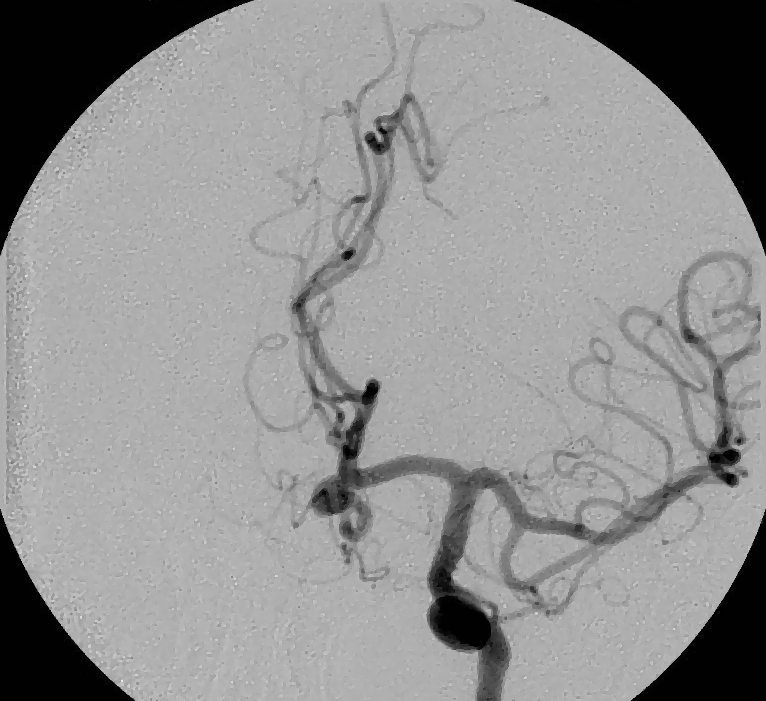} & 
		\includegraphics[width=1.53in]{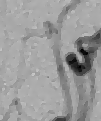}\\
	(d) corrupted&(e) TV$_{p}$-L$_1$ (ISNR = 19.10)&(f) zoom of (e)\\ 
		\includegraphics[width=2in]{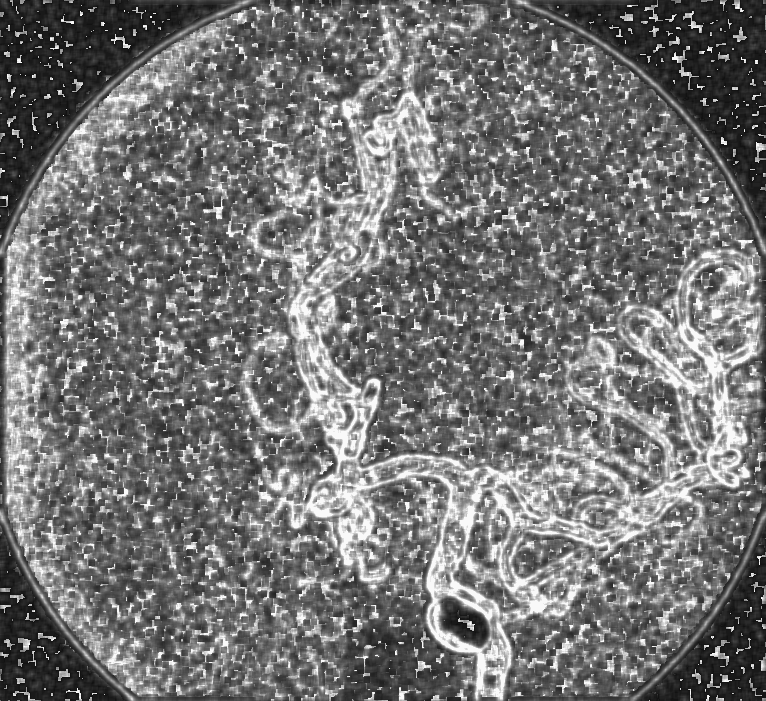} &
		\includegraphics[width=2in]{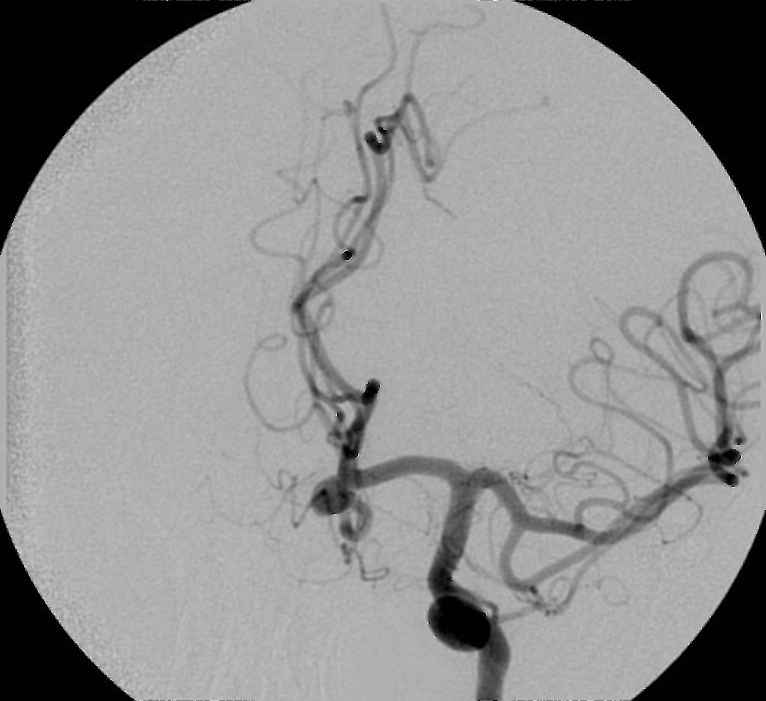} & 
		\includegraphics[width=1.53in]{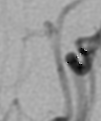}\\
			(g) $p$-map ($s=3$)&(h) TV$_{p}^{\mathrm{sv}}$-L$_1$ (ISNR = 21.14)&(i) zoom of (h)\\ 
		
		\includegraphics[width=2in]{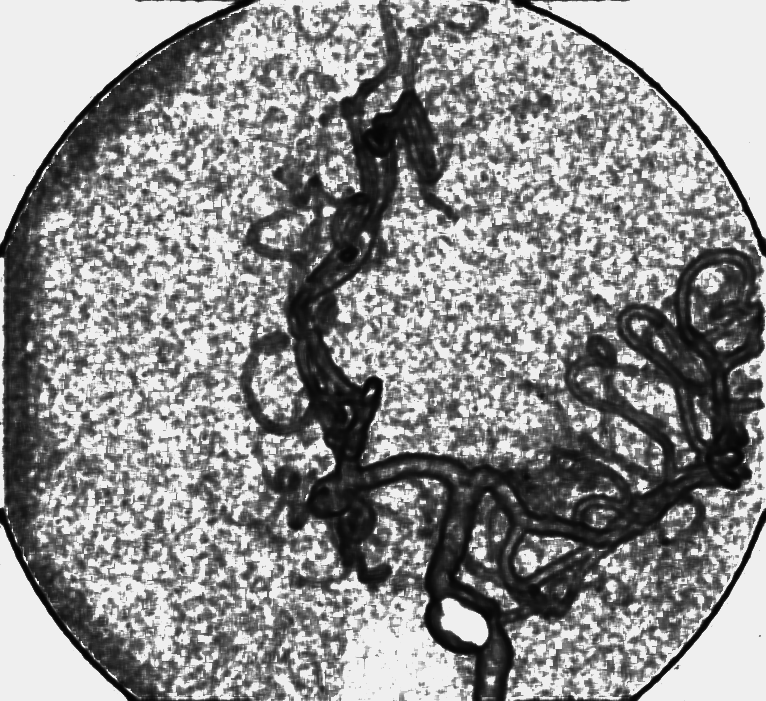} &
		\includegraphics[width=2in]{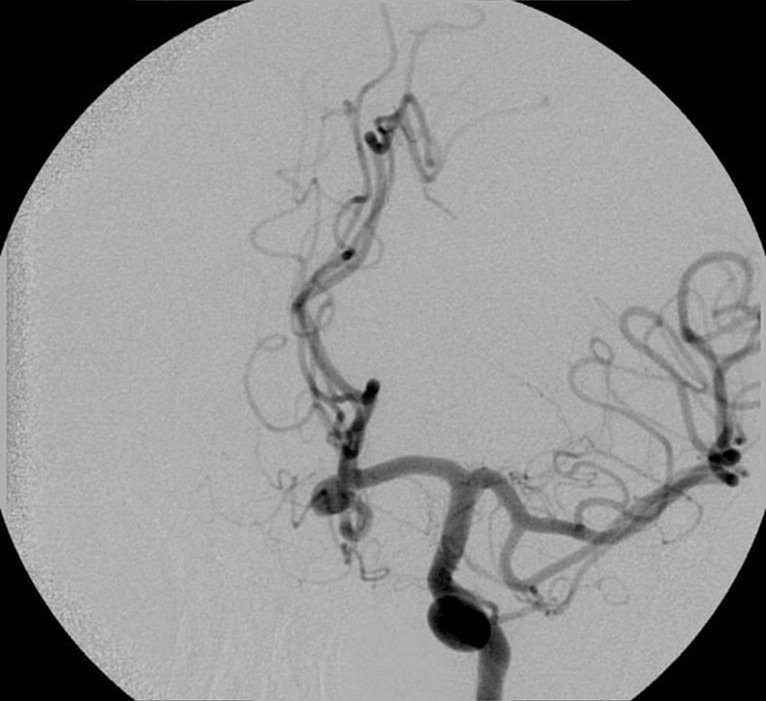} & 
		\includegraphics[width=1.53in]{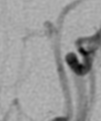}\\
	(l) $\alpha$-map ($s=3$)&(m) TV$_{p,\alpha}^{\mathrm{sv}}$-L$_1$ (ISNR = 24.47)&(n) zoom of (m)\\ 
\end{tabular}		
	\caption{Example 2 (SPN): visual restoration results for the test image \texttt{aneurism} corrupted by a $\gamma=0.1$ level noise.}
	\label{fig:aneu}
\end{figure}

First, for what concerns corruptions by SPN, in Figs. \ref{fig:lungs}, \ref{fig:eco}, \ref{fig:aneu} we report for the three considered test images the original and corrupted image together with the estimated $p,\alpha$-maps in the first column (with the size $s$ of the neighborhoods used for the $p,\alpha$-maps estimation reported in the captions), the restoration results, obtained by the four compared methods, in the second column (with the achieved ISNR values in the captions) and a zoomed detail of the restored images - green- bordered in Figs. \ref{fig:lungs} (a), \ref{fig:eco} (a), \ref{fig:aneu} (a) - in the last column.

The reported ISNR values as well as the visual inspection of the restored images and of the zoomed details strongly indicate how the proposed space-variant regularizer allows for higher quality restorations. In particular, it is worth remarking how, with respect to the space-variant TV$_p^{\mathrm{sv}}$ model, the additional degrees of freedom represented by the scale parameters $\alpha_i$ used in our proposal, yield a sufficient additional flexibility for avoiding unwanted spurious effects - see, e.g., spikes in Figs. \ref{fig:lungs} (i), \ref{fig:eco} (i), \ref{fig:aneu} (i). 

In the second part of this example, we consider the restoration of the same three medical test images corrupted by the same blur of parameters 
\verb|band=5| , \verb|sigma=1| and by a different impulsive noise, namely AWLN of level yielding BSNR=10. In Table \ref{tab:2} we report the ISNR values achieved by the compared methods and in Fig. \ref{fig:lapl} we show the original images, the corrupted images and the restored images by our model. 
The results in Table \ref{tab:2} confirm that, also in case of images corrupted by AWLN, the proposed TV$_{p,\alpha}^{\mathrm{sv}}$-L$_1$ model outperforms 
its competitors in terms of ISNR. Moreover, the restored images depicted in the last column of Fig. \ref{fig:lapl} provide further evidence of the good quality restorations achievable by our proposal.

\begin{figure}[tbh]
	\center
	\begin{tabular}{ccc}
		\includegraphics[width=2in]{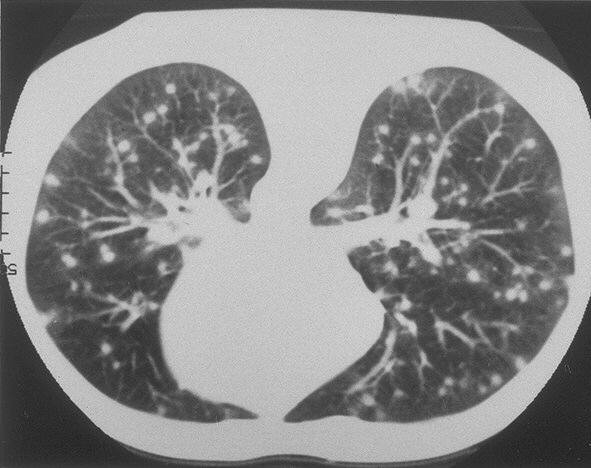} &
		\includegraphics[width=2.23in]{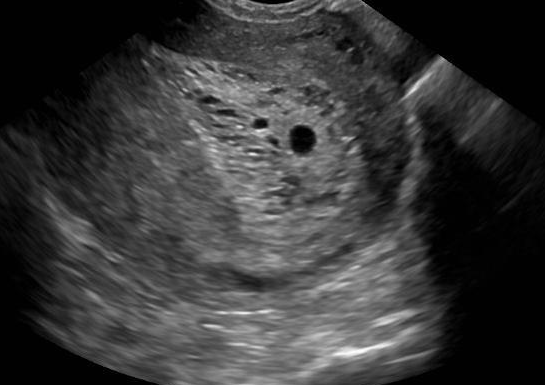} &
		\includegraphics[width=1.72in]{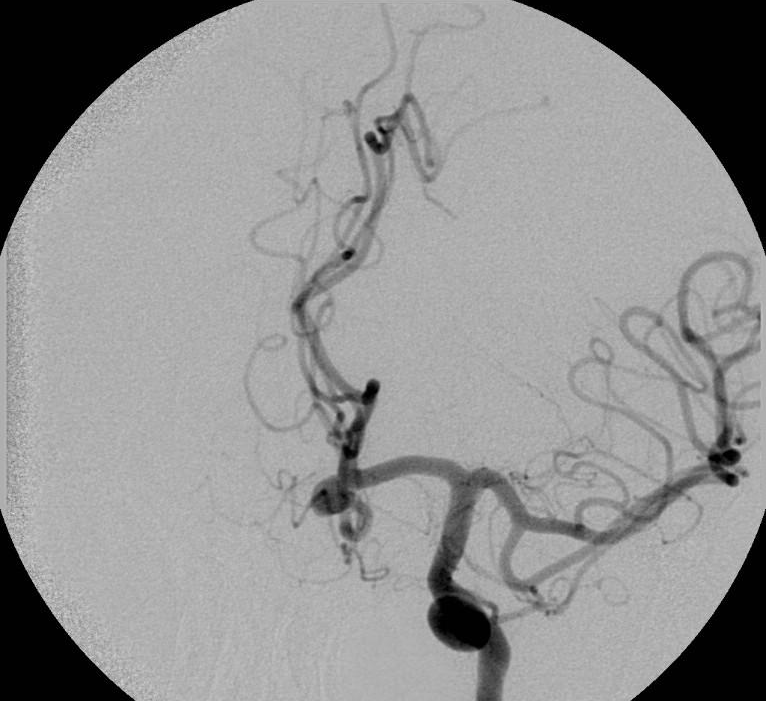}\\
		(a) original&(b) original&(c) original\\
		\includegraphics[width=2in]{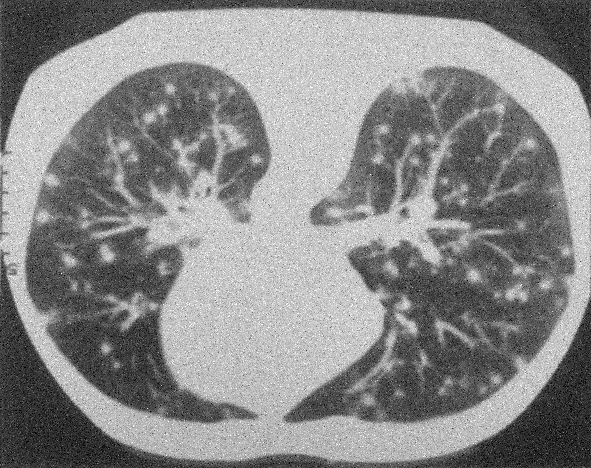} &
		\includegraphics[width=2.23in]{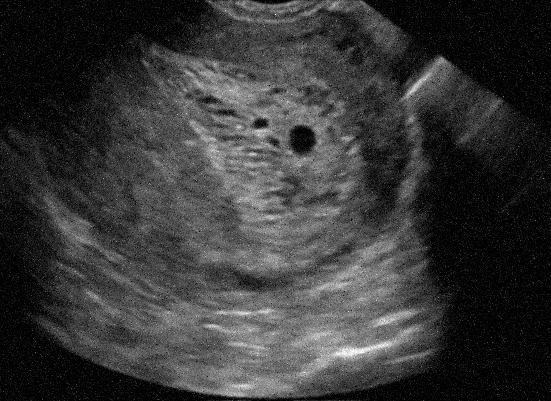} & 
		\includegraphics[width=1.72in]{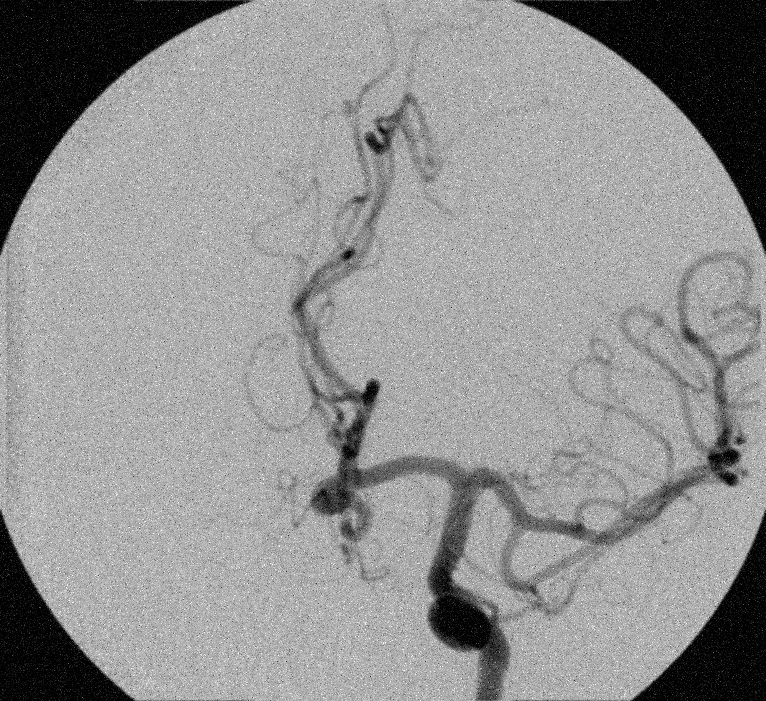}\\
		(d) corrupted (BSNR=10)&(e) corrupted (BSNR=10)&(f) corrupted (BSNR=10)\\ 
		\includegraphics[width=2in]{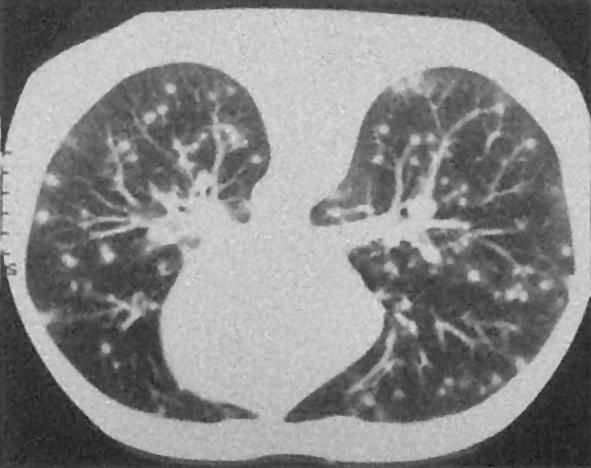} &
		\includegraphics[width=2.23in]{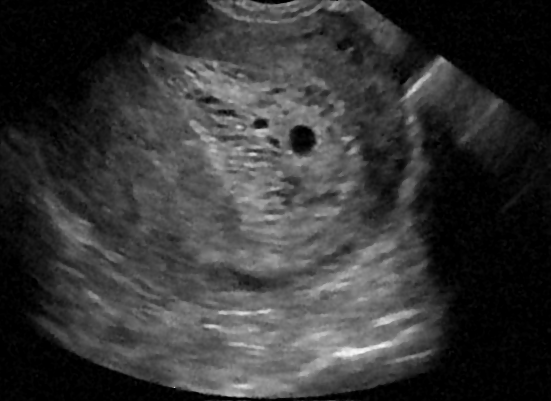} & 
		\includegraphics[width=1.72in]{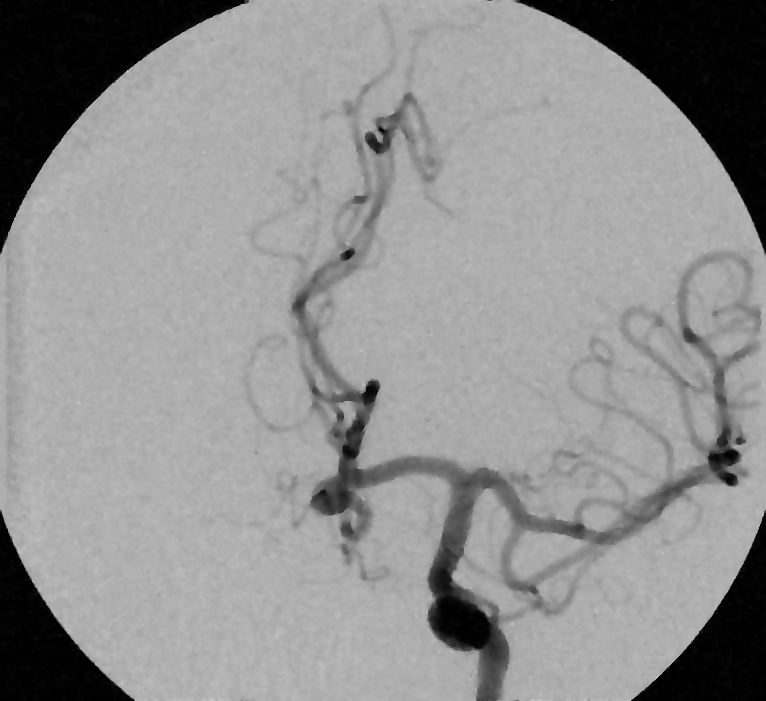}\\
		(g) TV$_{p,\alpha}^{\mathrm{sv}}$-L$_1$ &(h) TV$_{p,\alpha}^{\mathrm{sv}}$-L$_1$ &(i) TV$_{p,\alpha}^{\mathrm{sv}}$-L$_1$\\ 
	\end{tabular}		
	\caption{Example 2 (AWLN): visual restoration results.}
	\label{fig:lapl}
\end{figure}

\begin{table}
	\caption{Example 2 (AWLN): achieved ISNR values.}
	\label{tab:2}       
	%
	%
	\centering
	\begin{tabular}{lrrrr}
				\hline\noalign{\smallskip}

		 & TV-L$_1$      & TV$_{p}$-L$_1$ & TV$_p^{\mathrm{sv}}$-L$_1$ &TV$_{p,\alpha}^{\mathrm{sv}}$-L$_1$\\
		\noalign{\smallskip}\hline\noalign{\smallskip}
		\textbf{lungs}               & 6.20 & 6.80 & 7.30 & 7.85\\
	\textbf{ecography }             &5.93  & 6.40  & 7.88 & 8.32\\
		\textbf{aneurism}              & 9.10  & 9.44  & 10.13 & 10.70 \\
		
		\noalign{\smallskip}\hline\noalign{\smallskip}
	\end{tabular}
	
\end{table}


%
%
%
%
%

\section{Conclusions}
\label{sec:conc}
We presented a new space-variant regularizer for variational image restoration based on the assumption that the
gradient magnitudes of the target image distribute locally according to a half-Generalized Gaussian distribution. 
Thanks to the high number of free parameters involved in the regularizer and to the fact that such parameters can be automatically 
and robustly estimated from the observed image, our proposal exhibits a very high flexibility which potentially allows for 
an effective modeling of space-variant properties of images. 
By coupling the proposed regularizer with either the L$_1$ or L$_2$ fidelity terms, we tested our proposal on images corrupted by 
blur and different types of noise, namely AWGN, AWLN and SPN. The restored images, obtained by means of an efficient ADMM-based numerical algorithm, 
strongly indicate that the proposed regularizer holds the potential for achieving high quality restoration results for a wide range 
of target images characterized by different gradient distributions and for the different types of noise considered.\\

\textbf{Acknowledgments}: Research was supported by the ``National Group for Scientific Computation
(GNCS-INDAM)'' and by ex60 project by the University of Bologna ``Funds for selected research topics''.

\end{document}